\documentclass[a4paper,11pt]{article}
\pdfoutput=1 % if your are submitting a pdflatex
\usepackage{jinstpub}
\usepackage{graphicx}  % needed for figures
\usepackage{subfigure} % for side-by-side figures.
\usepackage{bm}        % for math
\usepackage{amsmath}   % for math
\usepackage{cancel} 
\usepackage{verbatim} 
\usepackage{hyperref}
\usepackage{seqsplit}

\title{Rapid-Cycling Synchrotron for Multi-Megawatt Proton Facility at Fermilab}
\author{J. Eldred,} 
\author{V. Lebedev and A. Valishev}
\affiliation{Fermi National Accelerator Laboratory \\Batavia, Illinois 60510 USA}
\emailAdd{jseldred@fnal.gov}

\abstract{The Fermilab accelerator complex delivers intense high-energy proton beams to a variety of fixed-target scientific programs, including a flagship long-baseline neutrino program. With the advent of the Deep Underground Neutrino Experiment (DUNE) and Long Baseline Neutrino Facility (LBNF) program there is strong motivation for a 2.4 MW beam power upgrade of the Fermilab proton facility. We show the Fermilab proton facility can achieve 2.4~MW with a new rapid-cycling synchrotron (RCS) to replace the Fermilab Booster and we provide a comprehensive technical analysis of the RCS-based facility design. Past design efforts and operational experience at the Fermilab Booster, J-PARC RCS, and Oak Ridge SNS are leveraged to provide strong empirical precedent for the design. We provide a parametric study of slip-stacking accumulation, RCS extraction energy, space-charge limits, beampipe aperture, eddy current heating, injection foil heating, and lattice optics. The 2.4 MW benchmark for the long baseline neutrino program is achieved independently of a previously proposed multi-GeV linac program, but we assess the impact the linac upgrade would have on RCS performance.}

\begin{document}
\maketitle

\section{Introduction}

\subsection{Overview}

\begin{figure}[htp]
\begin{centering}
\includegraphics[height=200pt]{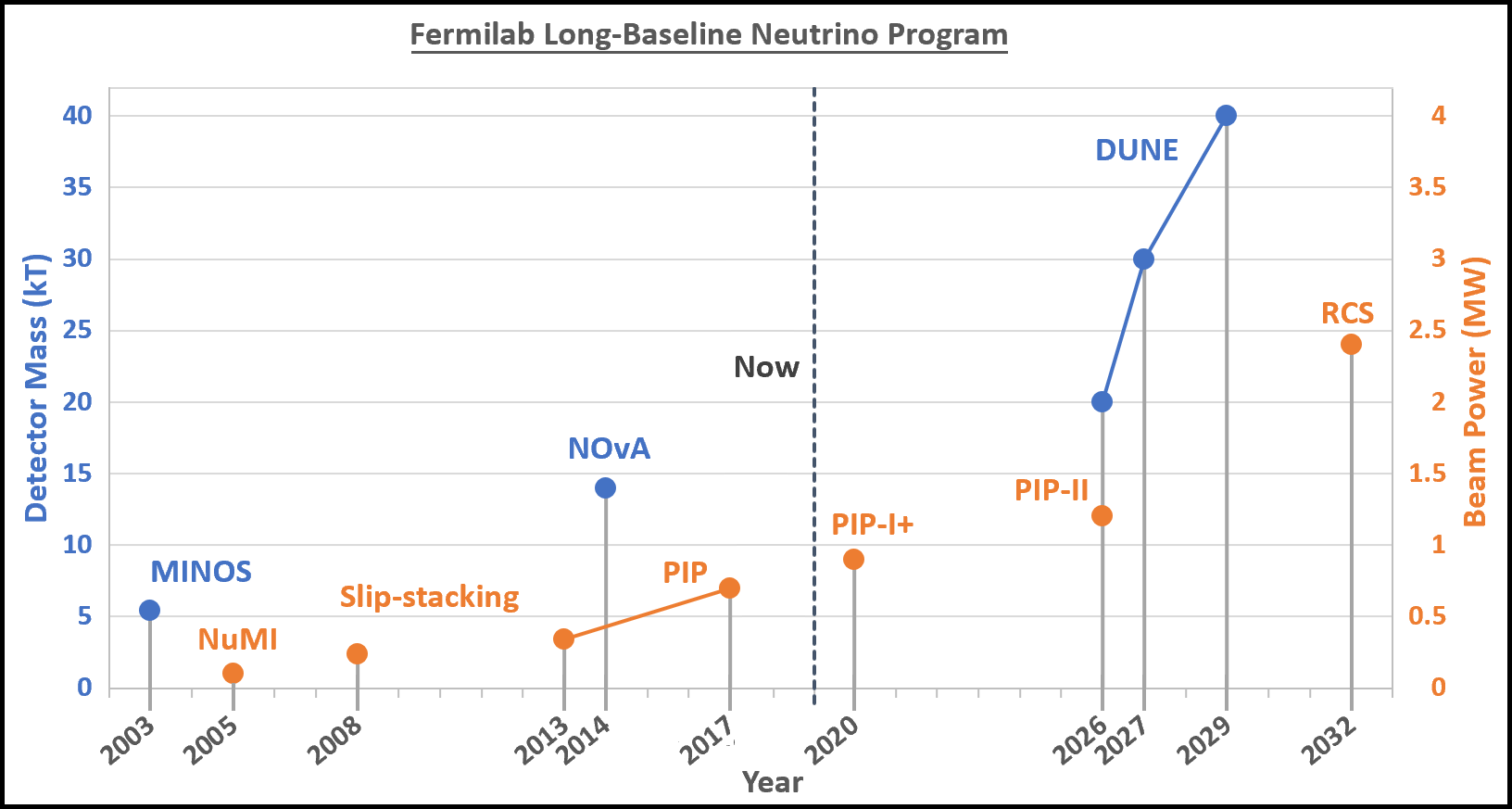}
 \caption{Past and Future milestones in Fermilab long-baseline neutrino program, as measured in detector mass and 120~GeV beam power.} 
  \label{History}
\end{centering}
\end{figure}

Whenever Fermilab has advanced the scale of its long-baseline neutrino detectors, it has been advantageous to increase proton power to the neutrino source commensurately.  Fig.~\ref{History} shows a timeline for detector~\cite{MINOS,NOvA,DUNEcdr2} and accelerator milestones~\cite{NuMI,Brown,AinsworthRR,Shiltsev,PIP2} of the Fermilab long-baseline neutrino program. The recent Deep Underground Neutrino Experiment (DUNE) Interim Design Report~\cite{DUNEidr} calls for an upgrade to the Fermilab proton complex of 2.4~MW beam power at 120~GeV by 2032. 

In this paper we perform a design study of the Fermilab proton complex to guarantee 2.4~MW Main Injector (MI) beam power without deviating from well-tested accelerator design principles. For the study we suppose that the Proton Improvement Plan II (PIP-II) upgrade will be completed (see \cite{PIP2}), but that we cannot rely on the Fermilab Booster or Recycler to be extended to 2.4~MW operation (see Section~\ref{SectionLimits}). For technical reasons, we pursue the design of a new rapid-cycling synchrotron (RCS) to replace the Fermilab Booster without a multi-GeV linac. We provide a parametric analysis of several critical design choices for an RCS-based proton facility and advance a specific scenario as a reference.

Our work is informed by recent operational experience at the Fermilab~\cite{AinsworthRR} and J-PARC~\cite{Hotchi}, the wider body of work on high-power ring design~\cite{TangRev,Wei,ChouRev}, and past iterations of Fermilab RCS design~\cite{PDriver,ICD2}.

We discuss the implications of the RCS for the broader Fermilab facility, including the high-power Main Injector and the experimental program for the new RCS beamline. We present several options for 3-5~MW future upgrades of the Fermilab proton complex that leverage the new RCS accelerator and anticipate emerging technology.

\subsection{Fermilab Proton Complex}

Broadly, the Fermilab proton accelerator complex is optimized for delivering intense high-energy proton beams to fixed-targets for the production of a variety of secondary and tertiary beams.

Presently, protons are accumulated in the Fermilab Booster via charge-stripping foil injection of 400-MeV H$^{-}$ ions from a conventionally conducting 30~mA linac. The 8-GeV proton pulses extracted from the Booster, called ``batches'', are sent to the short-baseline neutrino program~\cite{MicroBooNE,SBND}, the muon program~\cite{Stratakis,Mu2e,g2}, or to the Fermilab Recycler.

Up to seven Booster batches can be accommodated azimuthally in the Fermilab Recycler, but the Recycler accumulates twelve Booster batches using a technique known as slip-stacking. In slip-stacking, two beams are injected with a small momentum difference, each longitudinally focused by a corresponding RF cavity, in order to overlap the beams azimuthally (see \cite{EldredHSS}). Beam accumulated in the Recycler is transferred to the Main Injector and accelerated up to 120 GeV where it is either fast-extracted to the long-baseline neutrino program~\cite{NOvA} or slow-extracted to a fixed-target nuclear program~\cite{SeaQuest}.

\begin{figure}[htp]
\begin{centering}
\includegraphics[height=140pt]{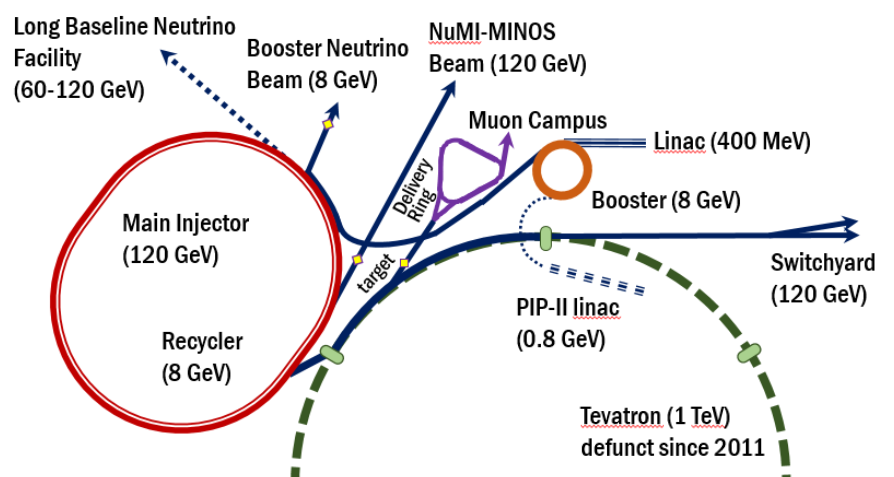}
 \caption{The Fermilab Proton Complex with past, current, and planned particle accelerator systems. Adapted from \cite{Shiltsev}.} 
  \label{PComplex}
\end{centering}
\end{figure}

\subsection{DUNE/LBNF}

The Deep Underground Neutrino Experiment (DUNE) is a massive liquid argon time-projection chamber that will be located deep underground at the Sanford Underground Research Facility in Lead, South Dakota. The Long-Baseline Neutrino Facility (LBNF) is a Fermilab Main Injector proton target facility that will provide an intense neutrino beam for long-baseline neutrino oscillation measurement at DUNE. DUNE and LBNF together constitute an international multi-decadal physics program for leading-edge neutrino science and proton decay studies~\cite{DUNEcdr2,DUNEidr}.

DUNE/LBNF is complementary to and competitive with other long-baseline neutrino experiments proposed on a similar timescale - Tokai to Hyper-Kamiokande~\cite{ICFA,T2K} and the European Spallation Source Neutrino Superbeam~\cite{Chakraborty}.

\begin{figure}[htp]
\begin{centering}
\includegraphics[height=140pt]{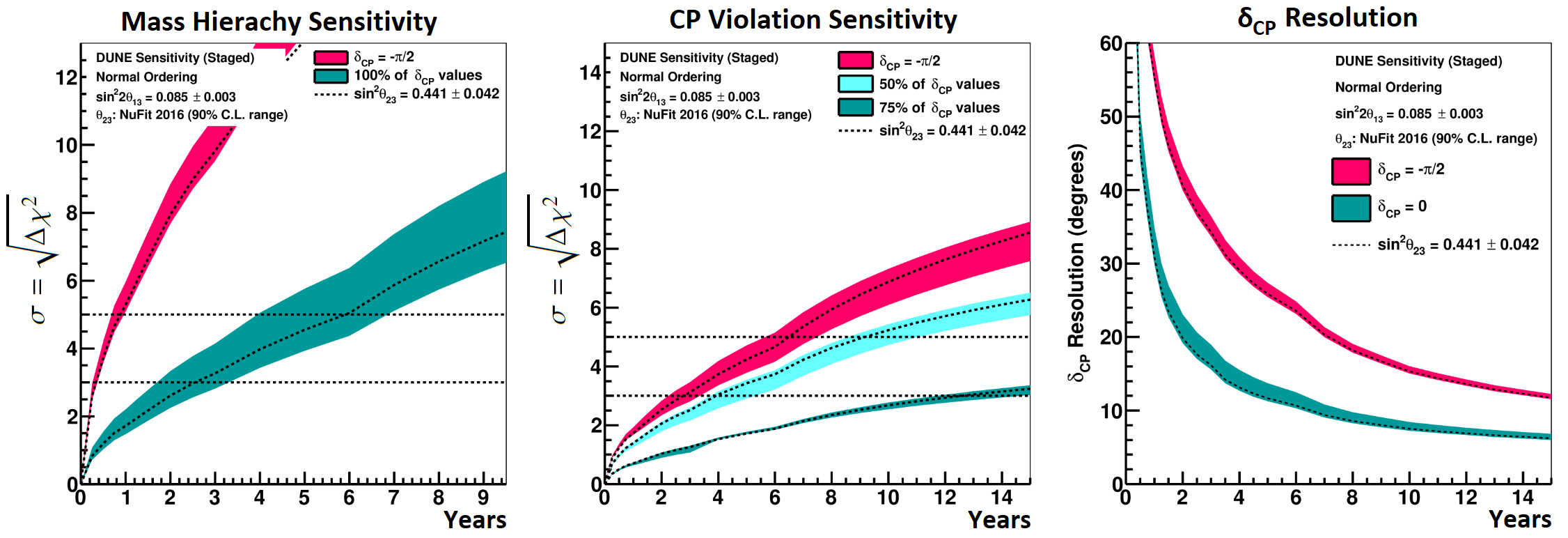}
 \caption{Anticipated DUNE physics results, adapted from \cite{DUNEidr}, with 2.4 MW upgrade at 6-year mark. (left) Sensitivity to determining neutrino mass hierarchy. (center) Sensitivity to determining nonzero CP-violating phase. (right) Resolution of CP-violating phase.} 
  \label{DUNE}
\end{centering}
\end{figure}

Currently the Fermilab proton complex delivers 700 kW to the long-baseline neutrino program by accelerating 52~$\times$~10$^{12}$ protons to 120 GeV in a Main Injector every 1.33 seconds. The DUNE Interim Design Report~\cite{DUNEidr} calls for an increase to 1.2 MW beam power at 120 GeV by 2026 and to 2.4 MW beam power at 120 GeV by 2032. Fig.~\ref{DUNE} shows the anticipated fundamental physics results with this staged power upgrade.

\subsection{1.2 MW Proton Facility with PIP-II}

The Proton Improvement Plan II (PIP-II)~\cite{PIP2} is a set of upgrades and improvements to the Fermilab accelerator complex to achieve 1.2 MW beam power at 120 GeV. In the summer 2018, the PIP-II project has received ``Critical Decision 1'' (CD1) recognition from the Department of Energy Office of Science.

PIP-II will feature several upgrades to improve Booster performance. The Booster batch intensity will need to increase 55\% from 4.2 $\times$ 10$^{12}$ to 6.5 $\times$ 10$^{12}$, while halving the loss rate in the Booster~\cite{Shiltsev}. PIP-II will include a new 2~mA CW-capable superconducting linac that will raise the injection energy into the Booster from 0.4~GeV to 0.8~GeV. The space-charge forces in the Booster will be reduced by the higher injection energy regardless of the higher beam intensity. The H$^{-}$ beam will also have a transverse rms emittance of less than 0.3~mm~mrad, allowing the Booster injection to be painted transversely. The new bunch-by-bunch MEBT chopper will allow the injected beam to be captured directly by the Booster RF, avoiding any losses associated with adiabatic bunching~\cite{Bhat}.

Along with Main Injector intensity increase, the 1.2~MW beam power will also require an 11\% reduction in the Main Injector ramp time from 1.33~s to 1.2~s. The Booster ramp rate will increase from 15~Hz to 20~Hz to facilitate slip-stacking~\cite{EldredThesis} and increase power to the 8~GeV beamline~\cite{MicroBooNE,SBND,Stratakis,Mu2e,g2}. Main Injector upgrades will also include a new set of pulsed quadrupoles for transition crossing as well as a RF power amplifier upgrade (see~Section~\ref{SectionMI}).

PIP-II also supports the proposed Mu2e-II program~\cite{Mu2eII} using the beam directly from the 0.8~GeV linac.

\subsection{Limits on Fermilab Booster and Recycler} \label{SectionLimits}

In order to achieve 2.4 MW power in the Main Injector, it will be necessary to replace the Fermilab Booster~\cite{PDriver,Shiltsev}. The Fermilab Booster is over 45 years old and faces limitations from its magnets and its RF alike. The dipoles serve as vacuum chambers without an interior beampipe, and the magnet laminations generate a significant impedance effect. 

Currently the impedance results in ~200 kV deceleration near transition crossing where the bunch length is shortest. After transition, this effect generates a large longitudinal emittance growth~\cite{Lebedev,PIP2}. Above PIP-II intensities, the expected beam loss would require the Booster to be rebuilt to accommodate a new beampipe in the dipoles. The uncontrolled particle loss rate would also have to be reduced considerably in a higher intensity Booster~\cite{Shiltsev} and the Booster suffers from significant field-errors that reduce the dynamic aperture~\cite{Alexahin,ValishevBooster}. 
It would also be challenging for the Fermilab Recycler to achieve double the PIP-II intensity and half the loss rate. The Recycler is a fixed-energy storage ring and would be unable to accept beam from an RCS at energies greater than 8-GeV. With or without slip-stacking, several betatron resonances would need to be completely suppressed to accommodate the increased space-charge tune-spread and chromaticity~\cite{Ainsworth}. Amplification of electron cloud in the combined function magnets of the Recycler has previously lead to a severe instability that delayed commissioning of the Recycler~\cite{Antipov}. At higher intensities the Recycler beampipe will require conditioning to mitigate the electron cloud instability, and this might constrain the duty-factor for operation at maximum intensity. The Recycler vertical beampipe diameter is 8\% smaller than the Main Injector vertical beampipe diameter~\cite{BrownCol}, which will constrain the emittance of the beam.

The current performance of the Recycler is described in \cite{AinsworthRR}. After PIP-II, we will become more confident in our projections of Recycler performance at high-intensity. At present, upgrade proposals should provide a path to 2.4 MW without use of the Fermilab Recycler. In Section~\ref{SectionUpgrades} we discuss replacing the Recycler with a higher energy storage ring to potentially achieve 3.6~MW beam power.

\subsection{RCS Proposals for Multi-MW Beam Power}

\begin{figure}[htp]
\begin{centering}
\includegraphics[height=180pt]{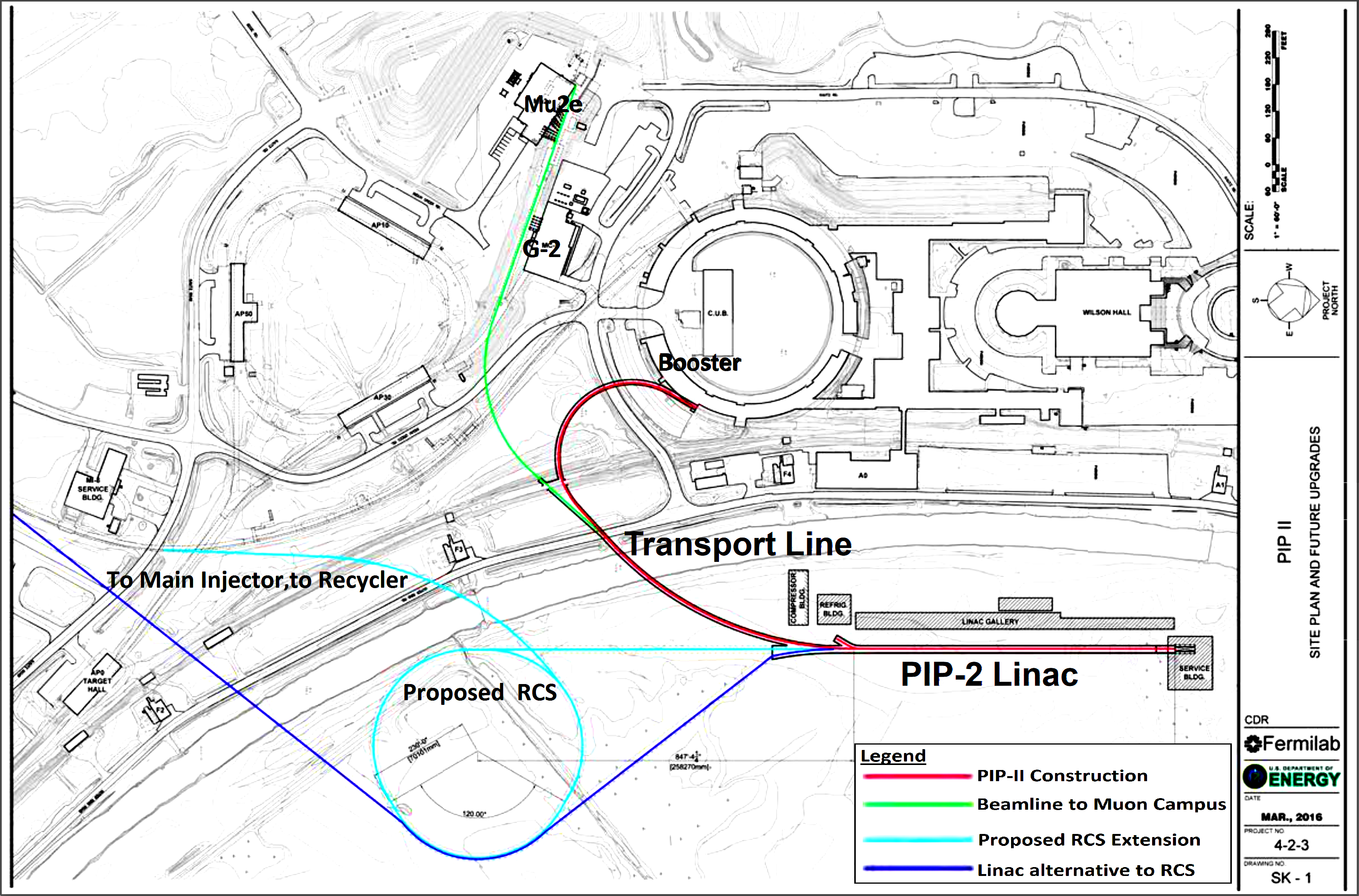}
 \caption{Site location for the proposed RCS, relative to the PIP-II linac, muon campus, and Main Injector. Adapted from \cite{Dixon}.} 
  \label{Siting}
\end{centering}
\end{figure}

Fig.~\ref{Siting} shows a diagram of the Fermilab complex with the footprint of the proposed RCS ring. The site location inside the Tevatron ring does not significantly constrain the size or shape of the RCS. 

The possibility of achieving 2~MW beam power in the Fermilab Main Injector by replacing the Booster with a new RCS was originally laid out in the 2003 Proton Driver Study II (PD2)~\cite{PDriver}. In 2010, the proposal received a major update as the Project X Initial Configuration Document 2 (ICD-2)~\cite{ICD2,Nagaitsev}. A comparison of major parameters is given in Table~\ref{PDtable}.

\begin{table}[htp]
\centering
\caption{Comparison between PIP-II Booster, PD2, ICD-2, and the RCS proposed in this paper}
\begin{tabular}{| l | l | l | l | l |}
\hline
~ & PIP-II & PD2 & ICD-2 & RCS \\
\hline
Beam Intensity & 6.5~$\times$~10$^{12}$ & 25~$\times$~10$^{12}$ & 27~$\times$~10$^{12}$ & 26~$\times$~10$^{12}$ \\
Circumference & 474~m & 474~m & 553~m & 600~m \\
Injection Energy & 0.8~GeV & 0.6~GeV & 2.0~GeV & 1.0~GeV \\
Extraction Energy & 8~GeV & 8~GeV & 8~GeV & 11~GeV \\
Transition Energy & 4.2~GeV & 13~GeV & 12~GeV & 13~GeV \\
Normalized Emittance & 15~mm~mrad & 40~mm~mrad & 25~mm~mrad & 24~mm~mrad \\
Ramp-Rate & 20~Hz & 15~Hz & 10~Hz & 15~Hz \\
Accumulation Ring & Recycler & Main Injector & Recycler & Main Injector \\
Accumulation Technique & Slip-stacking & Boxcar stacking & Boxcar stacking & Slip-stacking \\
MI Power & 1.2~MW & 1.8~MW & 2.2~MW & 2.4~MW \\
Available RCS Power & 80~kW & 350~kW & 200~kW & 440~kW \\
\hline
\end{tabular}
\label{PDtable}
\end{table}

The PD2 proposal uses a 95\% normalized emittance of 40 mm mrad, which is too large to be accommodated in the Main Injector without prohibitive losses. The ICD-2 reduces the emittance to 25 mm mrad and correspondingly increases the proposed injection energy to 2~GeV. The 2-GeV linac represents the larger scope of Project X, which is designed with the additional goal of supplying power to kaon and muon programs. The ICD-2 uses the Recycler for accumulation, whereas the PD2 stacks beam in the Main Injector directly.

A baseline RCS proposal presented in this paper has some important differences from the ICD-2 proposal. In Section~\ref{SectionLaslett} we show that the 2~GeV linac upgrade may not be necessary to achieve the required RCS intensity and consequently the linac-based experimental program can be considered independently. %In Section~\ref{Section11GeV} we discuss the possibility of a separate ring to slow-extract RCS batches for kaon and muon experiments.

Our RCS proposal does not use the Recycler to accumulate beam for the Main Injector; we instead reach 2.4~MW by using slip-stacking in the Main Injector to accumulate a total of nine batches in a 1.867~s cycle (Section~\ref{SectionPower}). We raise the RCS extraction energy to reduce the betatron tune-spread in the Main Injector and increase the normalized acceptance of the Main Injector. Our RCS design features a higher magnet strength and ramp rate, and consequently the beampipe heating from eddy currents must be managed by use of an Inconel beampipe (Section~\ref{SectionEddy}).

Like the PD2 and ICD-2, the Fermilab RCS lattice avoids transition-crossing and features dispersion-free insertions (Section~\ref{SectionLattice}). Unlike the PD2 and ICD-2, the RCS lattice is highly superperiodic to reduce the tune-spread per superperiod. The H$^{-}$ stripping foil injection, analyzed in Section~\ref{SectionFoil}, requires large betas in the injection straight and sufficient phase-advance to collimate particles scattered off the injection foil.

\section{Broad Design Features}  \label{SectionBroad}

\subsection{Slip-stacking Accumulation} \label{SectionSS}

Slip-stacking is an accumulation technique that nearly doubles the number of RCS batches that can be accumulated in the Main Injector, with a roughly 30\% increase in Main Injector cycle. Therefore to reach 2.4~MW beam power, a conventional boxcar-stacked RCS beam would require roughly 50\% more intensity than a slip-stacking RCS beam. The RCS with conventional beam accumulation would then require either a multi-GeV linac upgrade, an upgraded Main Injector ramp, or an advanced space-charge compensation technology (Section~\ref{SectionUpgrades} and Section~\ref{SectionIOTA}). To achieve 2.4~MW with a Recycler-like storage ring for beam accumulation, the required intensity is reduced by roughly 33\% for slip-stacking and by 14\% for conventional stacking. 

Slip-stacking accumulation was originally operationalized in the Main Injector, starting in 2004~\cite{Brown}. A robust beam-loading compensation system was developed to maintain stability of the the Main Injector RF cavities~\cite{DeySS,Seiya}. Slip-stacking accumulation in the Recycler began in 2015~\cite{AinsworthRR} to reduce Main Injector cycle time and slip-stacking in the Recycler is expected to continue through the PIP-II era.

\begin{figure}[htp]
\begin{centering}
\includegraphics[height=114pt]{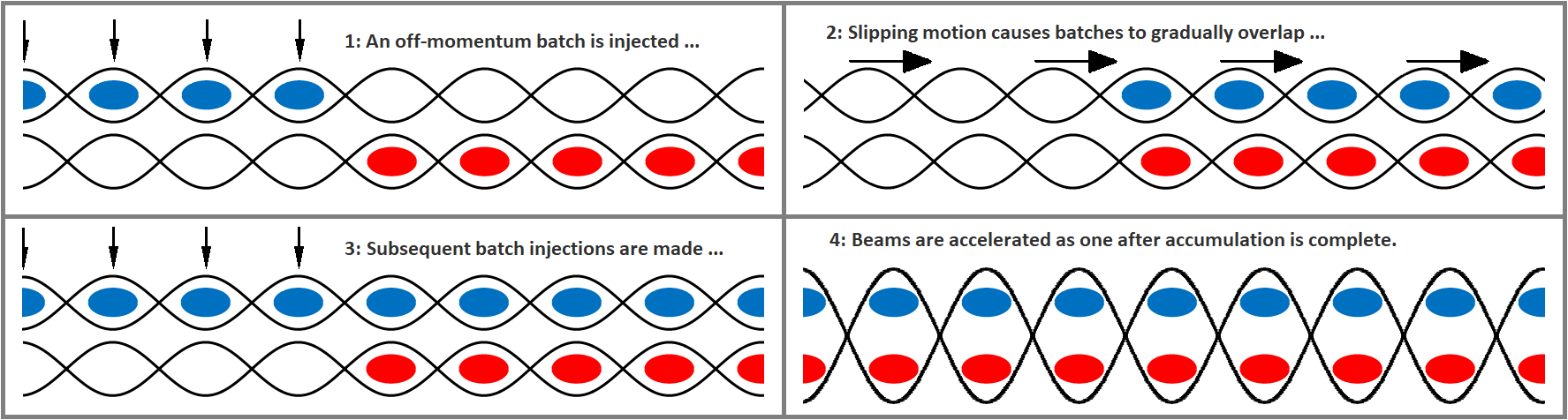}
 \caption{Diagram of slip-stacking accumulation. RF manipulation is used to gradually overlap the RCS batches in the Main Injector and thereby allow more batches to be accumulated in total.} 
  \label{SlipStacking}
\end{centering}
\end{figure}

Fig.~\ref{SlipStacking} illustrates that slip-stacking beams must slip one batch with respect to each other in time for the injection of each new batch. The RCS ramp-rate and circumference are connected to the RF frequency difference by the relation:
\begin{align} \nonumber
\Delta f = h_{\text{RCS}} f_{\text{RCS}} = \left( h_{\text{Booster}} \frac{C_{\text{RCS}}}{C_{\text{Booster}}} \right) f_{\text{RCS}}
\end{align}
where $h$ is the harmonic number, $C$ is the circumference and $f_{\text{RCS}}$ is the ramp-rate of the RCS. The difference in the frequency of the two RF cavities can be related to the difference in momentum of the two beams by the phase-slip factor $\eta$:
\begin{align} \nonumber
\Delta \delta = \frac{\Delta f}{f_{rev} h \eta}
\end{align}

The stability of the longitudinal dynamics of slip-stacking require the RF frequency difference to be much larger than the synchotron frequency, but a new technique using a second-harmonic RF cavity substantially relaxes that constraint~\cite{EldredHSS}. With the harmonic RF cavity, an RCS beam with a total longitudinal emittance of 0.2 eV$\cdot$s can be efficiently transported in the Main Injector if there is at least 750~Hz separation between the two RF systems.

The momentum span of the two slip-stacking beams will also need to accommodated within the momentum aperture of the Main Injector~\cite{EldredSS}. The momentum aperture ultimately depends on the Main Injector optics, but the RF frequency difference should be within 1680~Hz to be consistent with PIP-II requirements for the Main Injector (momentum span within 0.6\%). If the RF frequency difference corresponding to the RCS ramp rate is too high, it can be reduced by using partial batches or by alternating RCS cycles between the Main Injector and other beamlines.

A recent study of the Fermilab Recycler \cite{Ainsworth} characterized the betatron tune spread for intense slip-stacking beams. The tune-spread of the coincident slipping beams is comparable to the tune-spread of the beams after stacking. A new transverse damper system in the Recycler allows the coupled bunch instability to be mitigated while the beams are slip-stacking~\cite{Eddy}.

\subsection{RCS Circumference \& Ramp Rate} \label{SectionCircum}

\begin{figure}[htp]
\begin{centering}
\includegraphics[height=64pt]{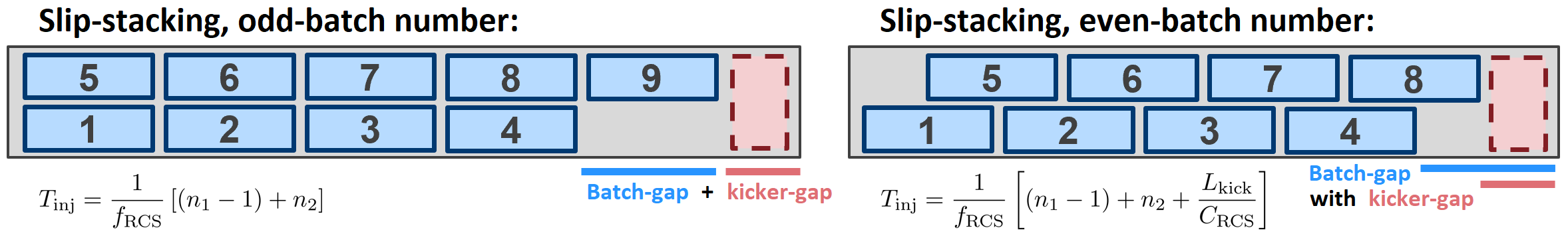}
 \caption{Accumulated batch structure for odd and even numbers of slip-stacked batches. Below, the corresponding accumulation time for efficiently slip-stacked RCS batches.} 
  \label{Batchgap}
\end{centering}
\end{figure}

The RCS circumference should allow for efficient stacking in the Main Injector. After the first $n_{1}$ batches are injected at one momentum, each subsequent off-momentum batch must be accommodated azimuthally before the batches can be overlapped in slip-stacking. After all batches are accumulated in the Main Injector an extraction kicker gap length $L_{\text{kick}}$ must remain for fast extraction at 120~GeV. For a generic RCS circumference, $n_{1}+n_{2}$ batches can be accommodated in the Main Injector via slip-stacking under the conditions:
\begin{align} \nonumber
C_{\text{RCS}} \cdot (n_{1}+1) & \leq C_{\text{MI}} \\ \nonumber
C_{\text{RCS}} \cdot \text{max}[n_{1},n_{2}]+L_{\text{kick}} & \leq C_{\text{MI}}
\end{align}
where $C$ is the machine circumference and $L_{\text{kick}} < C_{\text{RCS}}$. The Fermilab Main Injector circumference is 3319~m, and fast extraction from the Main Injector at 120 GeV requires a $\sim$1000~ns (300~m) kicker gap~\cite{Jensen}. Fig.~\ref{Batchgap} shows the batch structure after slip-stacking accumulation with the corresponding accumulation time.

The total Main Injector cycle time is the accumulation time $T_{\text{inj}}$ plus the Main Injector ramp time $T_{\text{ramp}}$ (1.2~s for the PIP-II 8~GeV to 120~GeV ramp), rounded up to the nearest RCS cycle $1/f_{\text{RCS}}$. The Main Injector ramp rate is discussed in Section~\ref{SectionMI} and Section~\ref{SectionUpgrades}.

The Main Injector power is a function of the Main Injector intensity, Main Injector energy, and cycle time $P_{MI} = e N_{\text{MI}}E_{\text{MI}}/(T_{\text{inj}}+T_{\text{ramp}})$. Expressing the Main Injector intensity can be given in terms of the RCS intensity $N_{\text{MI}} = (n_{1}+n_{2})N_{\text{RCS}}$, we find that the RCS intensity required to achieve 2.4~MW Main Injector power is just a function of the RCS ramp rate and circumference.

\begin{figure}[htp]
\begin{centering}
\includegraphics[height=180pt]{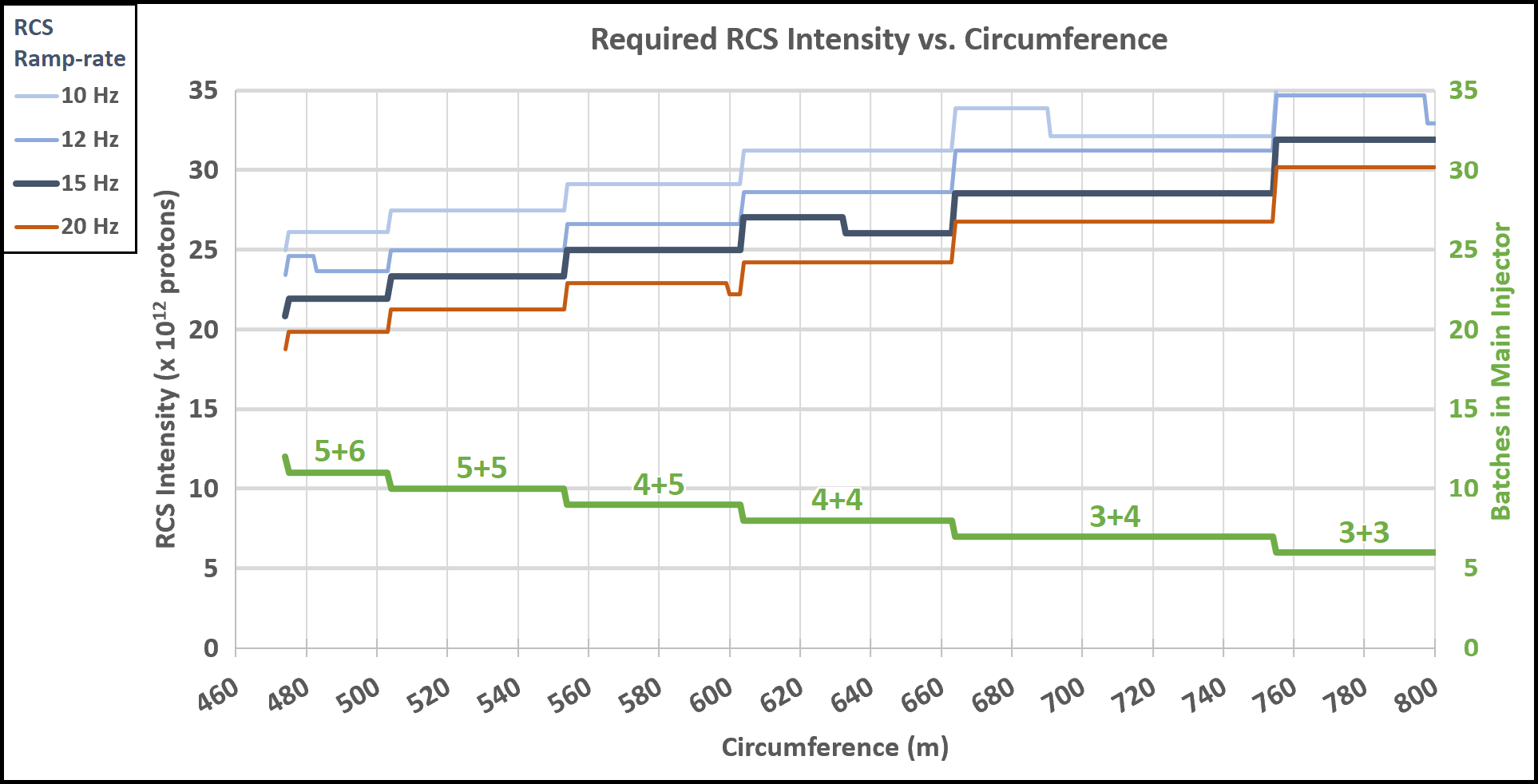}
 \caption{The number of RCS batches that can be accumulated via slip-stacking in the Main Injector is indicated by the green line (right-side axis). The RCS intensity required to achieve 2.4~MW in the Main Injector given that accumulation scheme is shown for several RCS ramp-rates (left-side axis).}
  \label{Batch}
\end{centering}
\end{figure}

Fig.~\ref{Batch} shows the required RCS intensity vs. circumference for selected ramp-rates. The model assumes an increase in RCS extraction energy with circumference that has a slight impact on Main Injector ramp time. Our analysis neglects the possibility of accumulating a partial RCS batch for the last injection, which arises from an inefficient choice of RCS circumference. 

\subsection{RCS Energy} \label{SectionEnergy}

For best performance of the RCS, the design emittance of the RCS should be increased until it is constrained by the acceptance of the Main Injector. At the same time, the beam will need to be accumulated in the Main Injector with minimal loss and at more than double PIP-II intensity. Consequently, the RCS extraction energy should be increased above 8~GeV to improve high-power performance of the Main Injector.

In particular, increasing the extraction energy of the RCS will strongly reduce the space-charge forces of the beam accumulating in the Main Injector. In Section~\ref{SectionLaslett} we argue that the Laslett tune-shift can be used as a generic proxy for space-charge forces in a ring. The geometric emittance is constrained by the Main Injector acceptance $\sim$4.3~mm~mrad, consequently the Main Injector Laslett tune-shift scales as a factor of $1/(\beta^{2} \gamma^{3})$. 

\begin{figure}[htp]
\begin{centering}
\includegraphics[height=180pt]{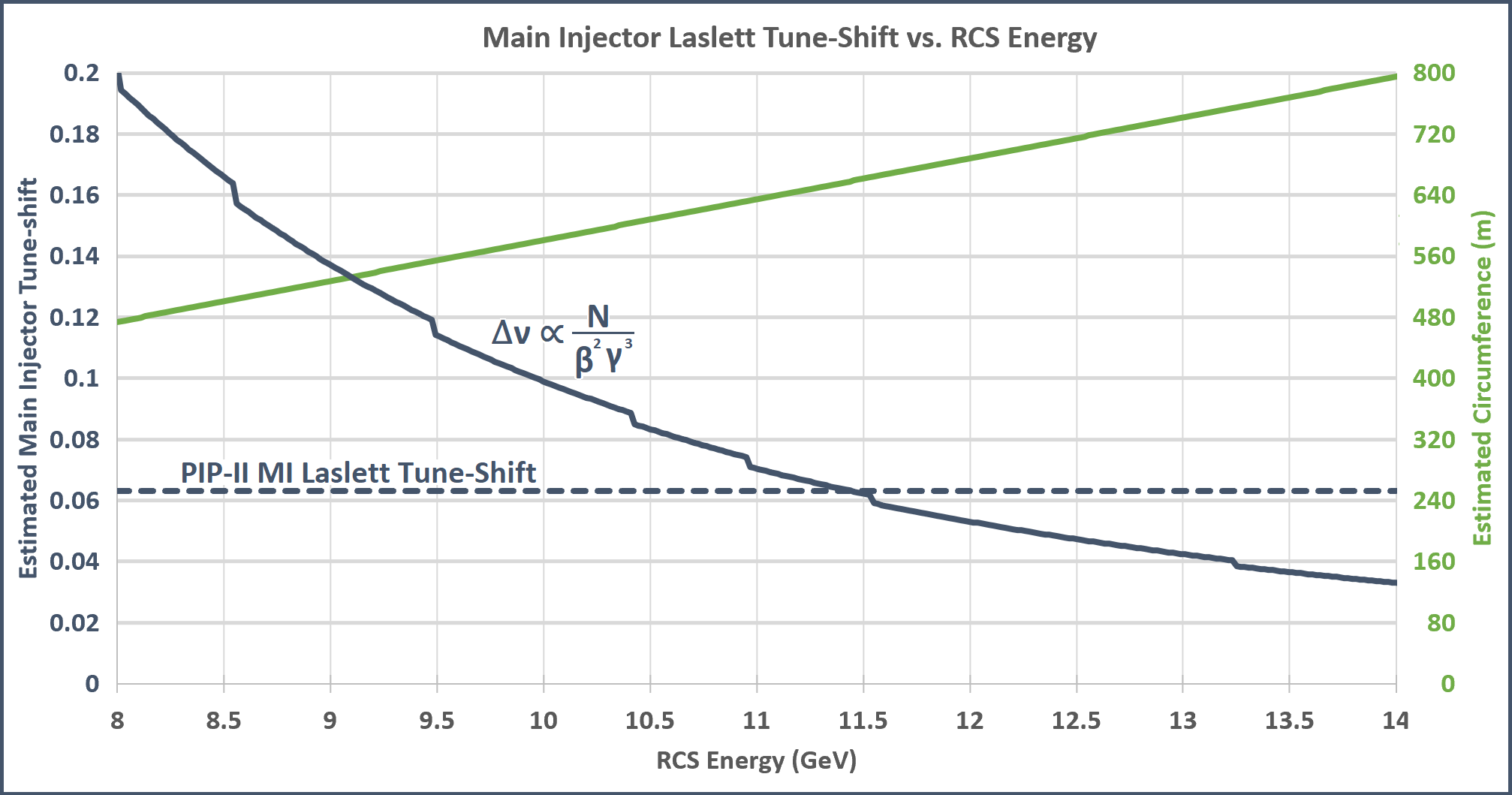}
 \caption{The estimate of RCS circumference as a function of RCS extraction energy is indicated by the green line (right-side axis). The estimated Main Injector Laslett tune-shift is indicated by the solid blue line (left-side axis). This Main Injector tune-shift can be compared to the PIP-II era Main Injector tune-shift indicated by the dotted blue line.} 
  \label{MIshift}
\end{centering}
\end{figure}

Fig.~\ref{MIshift} shows the Main Injector Laslett tune-shift as a function of RCS energy and a corresponding estimated RCS circumference. The RCS circumference is estimated by the RCS extraction energy by assuming the same integrated dipole field per length as the Booster. The required intensity of the Main Injector is also dependent on the accumulation scheme shown in Fig.~\ref{Batch} and we assume a 15~Hz RCS ramp-rate to estimate the Main Injector intensity as a function of circumference.

Table~\ref{Ctable} shows estimated parameters for several efficient choices of RCS circumference. The RCS circumference should be 600~m or 660~m to minimize space-charge forces in the Main Injector. We chose to consider an RCS with a 600~m circumference, but increase the integrated dipole per unit length to allow for an 11~GeV extraction energy. The integrated dipole per unit length is limited by eddy current heating beampipe heating (Section~\ref{SectionEddy}) and saturation effects of the ferrite dipole magnets ($\sim$1.4~T).

The Main Injector cycle time increases to 1.8~s and the Main Injector Laslett tune-shift is approximately 0.076 or 20\% larger than the PIP-II case. For comparison, the difference between the nominal vertical tune and the nearest third-order resonance in the Main Injector is 0.082. 

\begin{table}[htp]
\centering
\caption{Estimated Parameters for selected circumferences, assuming an RCS ramp rate of 15~Hz. Last row, a scenario achievable with greater integrated dipole field per length.}
\begin{tabular}{| l | c | c | c | c | c | c |}
\hline
Circum. & Batches & MI cycle & RCS Intensity & MI Intensity & RCS Energy & MI Tune-shift \\ \hline
474~m & 6+6 & 2.000~s & 21~$\times$~$10^{12}$ & 250~$\times$~$10^{12}$ & 8.0~GeV & 0.20 \\
500~m & 5+6 & 1.933~s & 22~$\times$~$10^{12}$ & 241~$\times$~$10^{12}$ & 8.5~GeV & 0.17 \\
550~m & 5+5 & 1.867~s & 23~$\times$~$10^{12}$ & 233~$\times$~$10^{12}$ & 9.4~GeV & 0.12 \\
600~m & 4+5 & 1.800~s & 25~$\times$~$10^{12}$ & 225~$\times$~$10^{12}$ & 10.4~GeV & 0.090 \\
660~m & 4+4 & 1.667~s & 26~$\times$~$10^{12}$ & 208~$\times$~$10^{12}$ & 11.5~GeV & 0.063 \\
\hline \hline
600~m & 4+5 & 1.800~s & 25~$\times$~$10^{12}$ & 225~$\times$~$10^{12}$ & 11.0~GeV & 0.076 \\
\hline
\end{tabular}
\label{Ctable}
\end{table}

The required RCS intensity rises with RCS circumference, but the normalized emittance at injection into the RCS scales with the extraction energy. Consequently, the increased RCS circumference will actually decrease space-charge forces at injection into the RCS by scaling the normalized emittance. If the RCS ramp-rate is fixed and the energy increases with circumference, the required RF acceleration will increase quadratically with circumference (see Section~\ref{SectionRF}). When considering immediate costs to construct the RCS, excessive circumference is clearly undesirable. For future upgrades however (Section~\ref{SectionUpgrades}), the RCS extraction energy may impact the ultimate intensity limits of the Main Injector.

\section{2.4~MW Proton Facility with RCS}  \label{SectionFacility}

\subsection{Facility Power} \label{SectionPower}

\begin{figure}[htp]
\begin{centering}
\includegraphics[height=180pt, width=360pt]{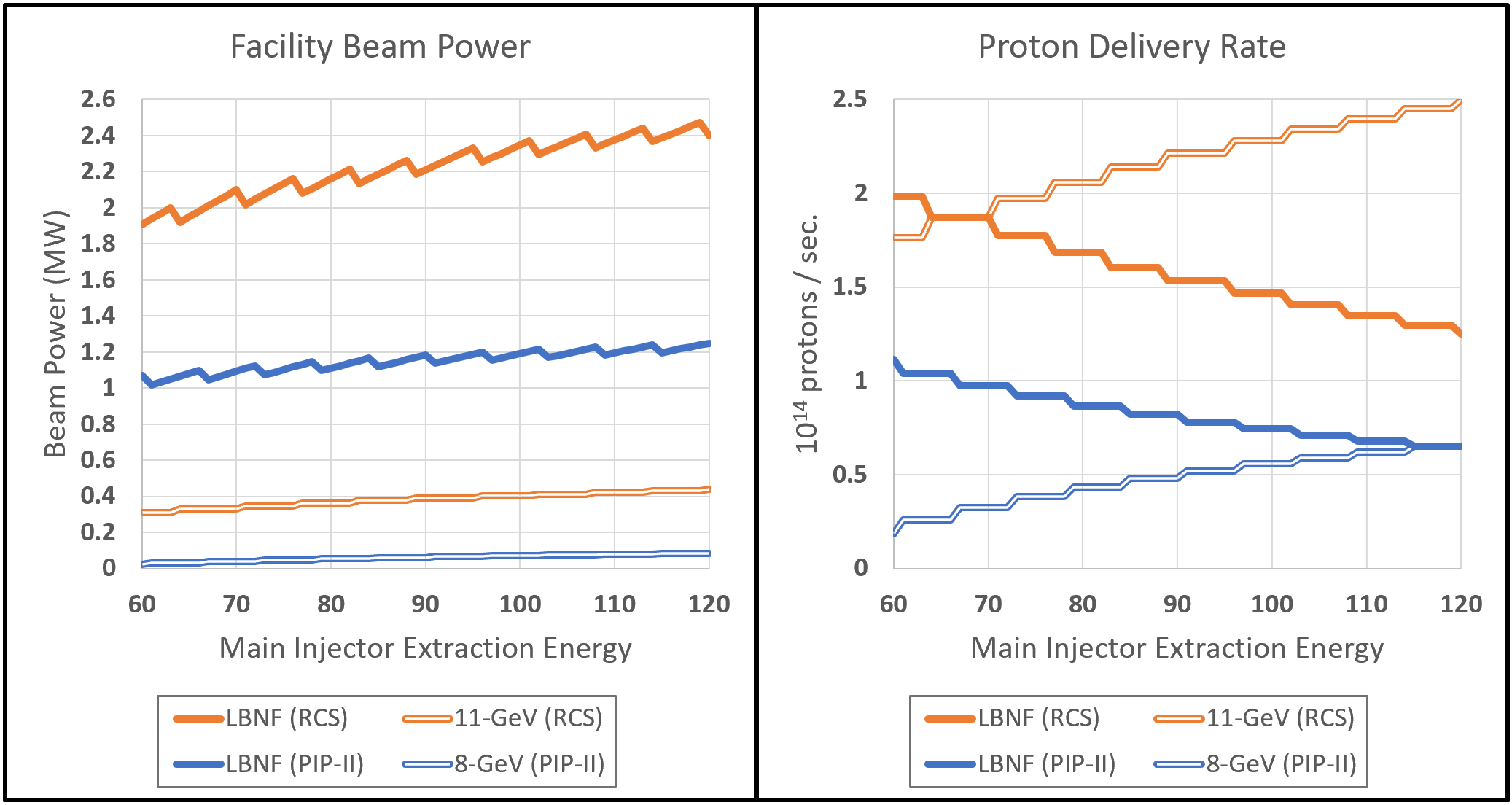}
 \caption{Proton beam power delivered to LBNF and 8-GeV/11-GeV beamlines, by extraction energy, during the PIP-II and RCS eras.} 
  \label{FacilityP}
\end{centering}
\end{figure}

The DUNE experimental program calls for the LBNF beamline to operate with a range of Main Injector energies from 60~GeV to 120~GeV~\cite{DUNEcdr2}. Figure~\ref{FacilityP} shows the power available at each beamline by Main Injector extraction energy and in comparison to PIP-II operation. When the beam energy delivered to LBNF beamline is reduced, the Main Injector ramp time is reduced and fewer RCS cycles are available for the 11-GeV beamline. 

\subsection{Main Injector Upgrades}  \label{SectionMI}

The 2002 Proton~Driver~Study Part~B~\cite{PDriver} addresses a wide array of topics needed to prepare the Main Injector for multi-MW operation. Slip-stacking is addressed in Section~\ref{SectionSS}. We discuss two major topics here - RF cavity power and Main Injector transition crossing.

Doubling the intensity of the Main Injector will have a significant beam-loading effect on its RF cavities. To provide the reactive power necessary to maintain beam stability, the Main Injector RF cavities will need to be replaced with a new design with larger power amplifiers (see \cite{Dey}).

%The replacement of the Main Injector RF cavities, combined with the new RCS, would be an ideal opportunity to change the fundamental RF frequency of the Fermilab proton complex. RF cavities with lower fundamental frequencies would tend to have a lower Q, larger aperture, and longer bunches.
%J-PARC Cavities citation?

The Main Injector crosses transition energy at 19~GeV which causes losses of less than 0.3\% of the beam~\cite{AinsworthGammaT,EldredThesis}. The accumulation of slip-stacked beam in the Main Injector increases the longitudinal emittance roughly be a factor of three and exacerbates transition crossing~\cite{Brown,EldredHSS}. With slip-stacking aided by a second-harmonic RF cavity, it may be possible to reduce the momentum separation between the beams just before capture and consequently reduce the longitudinal emittance. %In matching the RCS beam to the Main Injector, the longitudinal emittance can be varied operationally to strike a balance between minimizing space-charge forces during accumulation and minimizing losses from transition-crossing.

%At $\epsilon_{l} = 0.6 eV \cdot s$ and 225 $\times~10^{12}$ particles, the longitudinal space-charge defocusing at transition is about 15\% of the 4.4 MV RF focusing. 

To minimize losses at transition during PIP-II era operation of the Main Injector, a 0.5 unit $\gamma_{T}$-jump scheme based on pulsed triplets of quadrupoles will be implemented~\cite{PIP2}. With the $\gamma_{T}$-jump scheme and tunable longitudinal emittance, we do not anticipate any significant beam losses at transition crossing in the RCS era. An upgrade to a 1.0 unit $\gamma_{T}$-jump scheme is achievable with the pulsed quadrupoles~\cite{PDriver}, but that may require compensating the change to transverse optics~\cite{AinsworthGammaT}. The Fermilab Booster switches from negative chromaticity to positive chromaticity to reduce losses at transition, and this scheme could also be considered for the Main Injector.

During the PIP-II era, the Main Injector will ramp at 240 GeV/s and the Main Injector ramp time will be 1.2~s~\cite{PIP2}. This cycle time can be compared to an earlier study for the Proton Driver~\cite{PDriver} which found that the Main Injector magnets could be ramped over 1.165~s by operating at the performance limits of the existing magnet power supplies (a margin is recommended to ensure reliable operation in the summer months). The effect of an upgrade to ramp the Main Injector faster than 240 GeV/s is discussed in Section~\ref{SectionUpgrades}.

\subsection{LBNF Target}  \label{SectionLBNF}

The baseline LBNF design features a 1.2 MW segmented carbon target housed within a target hall rated for 2.4 MW~\cite{DUNEcdr3,LBNEv2}. Davenne et al.~\cite{Davenne} provides a detailed 2.4~MW target design based on helium-cooled segmented beryllium and optimized for production of neutrino beams from a 160~$\times~10^{12}$ proton pulse. The 2.4~MW scenario outlined in Section~\ref{SectionEnergy} requires a neutrino target rated for a higher pulse intensity, but with the addition of a 11-GeV storage ring (Section~\ref{SectionUpgrades}) the RCS-based proton facility could reach 2.4~MW beam power with only 150~$\times~10^{12}$ proton pulses.

In general, carbon and beryllium materials are preferred materials for high-intensity neutrino targets because of their low-Z value, radiation resistance, and thermal properties~\cite{LBNEv2,Simos}. The coolants under consideration are helium~\cite{Ahmad} and nitrogen~\cite{Tariq}. Ultimately the target and horn design must efficiently convert incident protons into a narrow neutrino beam with a desirable energy spectrum. Consequently the geometry of the target is heavily constrained and intimately connected with the focusing system for the secondary beam. Targetry, radiation damage, and thermal shock are all active areas of research for the development of intense neutrino targets.

\subsection{RCS Beamlines} \label{Section11GeV}

Thus far, the primary focus of our RCS design has been the beam power available to the Main Injector program. Nevertheless, the available beam power for 11~GeV beamline of the RCS represents more than a factor of five improvement over the PIP-II Booster.

For the 11-GeV program, the RCS ramp-rate is critical. The RCS ramp-rate of 15~Hz is constrained by the beampipe eddy current heating of Inconel 718 beampipe (Section~\ref{SectionEddy}). With a metalized ceramic beampipe design, the RCS ramp rate would only be limited by the total RF voltage that can be accommodated in the straight sections (Section~\ref{SectionRF}). At a ramp rate of 30~Hz, for example, the RCS could deliver 1.1 MW at 11 GeV (concurrent with 120 GeV operation at 2.4~MW).

%The Fermilab Booster could also be maintained for extraction to an 8-GeV beamline, delivering power up to 210 kW, but the cost of maintaining this older machine would need to be well-motivated.

Support for precision kaon and muon experiments was a central feature of the Project-X ICD-2 proposal~\cite{ICD2}. The Project-X 2-GeV superconducting CW linac constitutes a powerful proposal to support these physics programs and we present this as a staged upgrade option for our RCS proposal (Section~\ref{SectionUpgrades}).

There is also the option to support kaon and muon programs by slow-extraction from an 11-GeV storage ring. The duty-factor of the storage ring may have to be divided between the slow-extraction program and accumulation for the Main Injector program. The slow extraction mode of the storage ring would operate with stable optics near a third-order resonance~\cite{Pullia, Park} and control the spill rate by RF knockout~\cite{Nagaslaev}. Betas should be maximized at the extraction septa to minimize scattering losses~\cite{Nagaslaev2}.

%Extraction could continue over several batch injections if necessary, with a clearing-kicker removing the oldest batch every five injections. 

%Depending on the needs of the experiments, the RCS can extract some cycles at energies less than 11~GeV. If the RCS was operated with an 8~GeV beamline, the Fermilab Debunched could be used or the Fermilab Recycler could be retrofitted as a slow-extraction ring.

%A more detailed study is still needed to determine the feasibility of a slow-extraction Recycler, including changes to Recycler optics and shielding for high-loss regions of the tunnel shared with the Main Injector. An estimate of slow-extraction power can be obtained from existing operations in the Recycler. Current Recycler losses, at the few percent level, are comparable to septa losses from third-order extraction~\cite{Nagaslaev2}. This comparison suggests that Recycler may support slow-extraction of about three RCS batches of $26~\times~10^{12}$ protons every 1.8~s or 53~kW. However, a similar benchmark exercise for the current SeaQuest loss budget~\cite{Johnstone} suggests only two RCS batches per cycle. 

\section{RCS Design} \label{SectionDesign}

\subsection{RCS Emittance}

The emittance of the RCS beam will be ultimately constrained by the Main Injector acceptance at the RCS extraction energy. The normalized emittance corresponding to that constraint will then determine the required acceptance of the RCS at the injection energy. The space-charge forces at injection and peak temperature of the injection foil are also directly impacted.

The Main Injector total acceptance is a least 40~mm~mrad normalized at 8~GeV~\cite{Brown} and currently accommodates a Booster beam with a 95\% normalized emittance up to about 18~mm~mrad at extraction~\cite{Yang1,Yang2}. With the adiabatic damping from 8~GeV to 11~GeV, this corresponds to a beam with a 95\% emittance of 24 mm~mrad. This factor of 2.2 between the 95\% beam emittance and total ring acceptance can be compared to the J-PARC Main Ring where the factor is only 1.9~\cite{Saha,Harada}. Like the J-PARC facility, any beam halo generated during the RCS ramp will need to be collimated in the transfer line.

\subsection{Laslett Tune-shift} \label{SectionLaslett}

The Laslett tune-shift, the averaged tune shift by the direct space-charge in a ring, can be expressed
\begin{align} \label{Laslett}
\Delta \nu = \frac{N r_{0}}{2\pi \epsilon_{n} \beta \gamma^{2}} FB = \frac{N r_{0}}{2\pi \epsilon_{g} \beta^{2} \gamma^{3}} FB
\end{align}
where $N$ is the ring intensity, $r_{0}$ is the classical proton radius, $\epsilon_{n/g}$ is the normalized/geometric emittance. $F$ is the transverse form factor, which is 3 if $\epsilon_{n}$ is 95\% Gaussian emittance. $B$ is the bunching factor, the peak current over the average current~\cite{Prebys}. 

The true tune-shift due to space-charge should incorporate beampipe image charges~\cite{Chao}, but we focus on the tune-shift from the self-field because it is used as a generic proxy for the severity of space-charge forces in rapid-cycling synchrotrons and accumulator rings~\cite{ShiltsevNote,TangRev,Prebys}.

The space-charge tune-shift is not only a metric for charge density adjusted for relativistic effects, but is also an upper bound on the space-charge tune-spread. The betatron tune-spread is limited by the presences of betatron resonance lines which can be driven by magnet imperfections or by the space-charge forces themselves. Particles crossing resonance lines may be lost, and these particles must be collimated efficiently in order to keep within loss limits.

For a lattice with $k$ perfectly identical superperiods, only the $nQ_{x} + mQ_{y} = kp$ resonances are possible. The Fermilab Booster does not gain the full benefit of its 24-fold design symmetry~\cite{Alexahin,ValishevBooster}, but modern optics can correct beta-beating to better than 5\%~\cite{Malina}. The RCS ring design should be optimized for superperiodicity, and in Section~\ref{SectionLattice} we give an example with up to eight-fold symmetry. The expected performance of the RCS should be compared not just on the basis of the global tune-shift, but the tune-shift per superperiod.

The J-PARC RCS has demonstrated successful operation with a Laslett tune-shift of 0.3 and a corresponding tune-shift per superperiod of 0.1~\cite{Hotchi}. We propose to operate the Fermilab RCS with the higher tune-shift of 0.35, but a smaller tune-shift per superperiod of 0.044 .

The $26~\times~10^{12}$ proton intensity described in Section~\ref{SectionBroad} can be achieved with an injection energy of 1.0~GeV, 95\% normalized emittance $\epsilon_{N}$ of 24~mm~mrad, FB of $\sim 5$, and maximum Laslett tune of 0.35. The transverse form factor and bunching factor are consistent with the PIP-II Booster beam which is painted transversely and longitudinally~\cite{PIP2}. That scenario calls for a modest upgrade of the PIP-II linac to 1.0~GeV, which can be accommodated within the linac enclosure already planned~\cite{PIP2}. If the linac energy were to remain at 0.8~GeV, the corresponding Laslett tune shift is 0.44 .

\begin{table}[htp]
\centering
\caption{RCS Parameters}
\begin{tabular}{| l | l | l |}
\hline
Intensity & N & $26~\times~10^{12}$ protons \\
Circumference & C & 600 m \\
Ramp Rate & $f_{ac}$ & 15 Hz \\
\hline
Injection Energy & E$_{i}$ & 1.0 GeV \\ 
Extraction Energy & E$_{f}$ & 11 GeV \\ 
95\% Norm. Emit. & $\epsilon_{n}$ & 24 mm mrad \\
Laslett Tune Shift & $\Delta \nu$ & 0.35 \\
\hline
\end{tabular}
\label{RingParam}
\end{table}

J-PARC uses 2nd-harmonic RF cavities with total RF voltage corresponding to roughly one third of the main RF~\cite{Tamura,Hotchi2}, to improve the bunching factor and beam uniformity of the beam at injection. A similar investment in 2nd-harmonic RF cavities in the Fermilab RCS would reduce the Laslett tune-shift by about a factor of 1.5, for a tune-shift of $\sim 0.24$ .

%It may also be possible to achieve unusually large Laslett tune-shift with the use of innovative technology such as integrable optics or space-charge compensation (see Section~\ref{SectionIOTA}).

\subsection{Beampipe Aperture \& Eddy Currents} \label{SectionEddy}

For a beam with a 95\% normalized emittance $\epsilon_{n}$ the $k \sigma$ beampipe aperture for a round beampipe can be approximately given by
\begin{align} \label{a(k)}
a(k) = \frac{k}{1.96} \left[ \text{max}[\beta_{x},\beta_{y}] \frac{\epsilon_{n}}{\gamma \beta}  \right]^{1/2} + \text{max}[D_{x},D_{y}]~\delta_{\text{max}}
\end{align}

Following the precedent of the J-PARC RCS~\cite{Hotchi}, we set the collimator acceptance to be the 3$\sigma$ aperture of the beam and the total ring acceptance to be a factor of 1.5 greater than the collimator acceptance. This 3.6$\sigma$ total aperture can be compared to the Fermilab Booster where the total aperture is just under 3$\sigma$ in the vertical plane~\cite{SeiyaAlign}.

For a beam with a 24~mm~mrad normalized emittance at 1.0 GeV and a maximum beta function of 13~m, the 3.6$\sigma$ aperture is only a 2.45~cm radius. Incorporating a maximum dispersion of 1.2 m and a $\delta_{\text{max}} = 0.5~\times~10^{-3}$, a 2.51~cm radius aperture is required.

Aperture is a consideration in magnet cost, because the physical size of accelerator magnets generally scale with aperture squared (transversely, to capture the return flux) or cubed (longitudinally, to maintain the same integrated field). Consequently the lattice should be optimized to minimize beta function especially within high-field magnets.

Aperture is also a critical factor in minimized heating from the eddy currents generated by the magnet ramp. The eddy current power dissipated in the beampipe per surface area follows the scaling:
\begin{align}
\frac{dP}{d\Omega} \propto \sigma_{0} d a^{2} f_{\text{RCS}}^{2} B_{\text{AC}}^{2} 
\end{align}
where $\sigma_{0}$ is electrical conductivity, $d$ is pipe thickness, $a$ is beampipe radius, $f_{\text{RCS}}$ is the magnet ramp rate, and $B_{\text{AC}}$ is the amplitude of the magnetic field ramp~\cite{Nagaitsev}. 
For a beampipe surface in convective contact with relatively still air, a conservative estimate of the heat transfer coefficient is $10^{-3}~\text{W}~\text{cm}^{-2}~\text{K}^{-1}$. To avoid the need for active cooling of the beampipes, the change in temperature should be limited to $\sim 20 \text{K}$ and the corresponding power per surface area should be restricted to $\sim 20~\times~10^{-3}~\text{W}~\text{cm}^{-2}$.

The ideal beampipe material to minimize the eddy current heating should have high resistivity and ultimate tensile strength. Inconel 718 is an alloy primarily composed of Nickel, Iron, and Chromium that outperforms stainless steel 316 L on these measures, as shown in Table~\ref{InconelT}. By constructing a thinner and more resistive beampipe made from Inconel 718, the eddy current heating can be reduced by up to a factor of 4~\cite{Leibfritz}.
%\cite{Leibfritz,SpecialM,UnitedS}

\begin{table}[htp]
\centering
\caption{Beampipe Materials}
\begin{tabular}{| l | l | l |}
\hline
Parameter & Steel 316L & Inconel 718 \\
\hline
Resistivity & 73 $\mu \Omega$ cm$^{-1}$ & 125 $\mu \Omega$ cm$^{-1}$ \\
Tensile Strength & 80 ksi & 180 ksi \\
\hline
\end{tabular}
\label{InconelT}
\end{table}

A mechanical stress study~\cite{Tang} found that a 10~cm~$\times$~15~cm elliptical Inconel 718 beampipe could be constructed with a 0.20 mm wall thickness and 0.46 mm spiral supports. In this paper we propose a more conservative 5.02 cm diameter beampipe with a 0.45 mm wall thickness. The planned $\gamma_T$-jump for the PIP-II Main Injector will use Inconel beampipe inside the pulsed quads, providing Fermilab operational experience with the beampipe material.

The eddy current heating is directly related to the RCS lattice design. Eq.~\ref{a(k)} shows that the square of the beampipe radius $a^{2}$ is proportional to the maximum beta-function $\text{max}[\beta_{x},\beta_{y}]$ (neglecting dispersion as a small factor). The dipole field $B_{AC}$ is inversely proportional to bending radius $\rho$. We therefore introduce the eddy figure of merit $\chi$ that allows the RCS lattice to be directly optimized for its impact on eddy current heating:
\begin{align} \nonumber
B_{AC} &= \frac{1}{2} \frac{p_{f} - p_{i}}{e \rho} \\
\frac{dP}{d\Omega} &\propto \sigma_{0} d \epsilon_{n} f_{\text{RCS}}^{2} (p_{f} - p_{i})^{2} \chi \\
\chi &= \text{max}[\beta_{x},\beta_{y}]~\rho^{-2}
\end{align}

Eddy current heating places a constraint on the RCS ramp rate $f_{\text{RCS}}$, peak momentum $p_{f}$, and beam emittance $\epsilon_{n}$, but optimizing the eddy figure of merit $\chi$ can relax that constraint. The example lattice shown in Section~\ref{SectionLattice} demonstrates a value of $\chi = 0.010$, by using tighter betas and stronger magnets in the bending arc and then using wider aperture and weaker magnets in the straight sections.

With an eddy figure of merit of $\chi = 0.012$ or less and an Inconel~718 beampipe, the beampipe temperature increase from the RCS ramp can be kept below 20 degree without forced air cooling, while maintaining a 15 Hz ramp rate, 11~GeV extraction energy, and normalized ring acceptance of 79~mm~mrad. Minimizing the eddy figure of merit $\chi$ further generally comes at the cost of increased lattice circumference.

Although the Fermilab Booster does not have a beampipe in its dipoles magnets, we can calculate the eddy figure of merit for the Booster lattice to be $\chi = 0.017$. The eddy currents are very significant for the J-PARC RCS, where the eddy figure of merit is $\chi = 0.22$ and the beam emittance is an order of magnitude larger than the Fermilab Booster.

The J-PARC RCS manages its powerful eddy currents with titanium-flanged ceramic {beampipe \cite{Kinsho}}. A thin coating of a low secondary electron yield (SEY) material, such as titanium nitride, must be applied to the interior of the beampipe because of the large SEY of alumina ceramic. To avoid powerful impedance effects on the beam, the outside of the ceramic beampipe uses copper strips bridged with capacitors.

If the Fermilab RCS were to adopt a similar beampipe design, it would effectively remove the eddy current heating limit on the RCS ramp rate. The primary beneficiary of an increased RCS ramp rate would be the 11~GeV beamline. Slip-stacking accumulation rate in the Main Injector would still be limited by momentum difference of the slip-stacking beams (see Section~\ref{SectionSS}). 

Impedances of the beampipe are expected to be an order of magnitude weaker than the Booster impedances, which are dominated by the laminated dipoles~\cite{Lebedev2}. The RCS expects weaker collective instabilities despite its higher charge-density, and consequently can manage instabilities with a combination of chromaticity and bunch-by-bunch dampers. Electron cloud instabilities, such as the one that impacted the Fermilab Recycler~\cite{Antipov}, would be mitigated by a low-SEY coating beampipe coating such as titanium nitride or diamond-like carbon~\cite{Backfish,EldredEC}.

\subsection{RCS Magnet Ramp and RF Acceleration} \label{SectionRF}

Above 10~Hz, the RCS magnet power supplies should use resonant circuit power converters~\cite{Tang}, in which the ramp of the magnetic field follows a sinesoidal shape distorted by dynamic inductance. Higher harmonics of the power supply ramp frequency can be controlled to preserve the sinesoidal ramp shape of the magnetic-field~\cite{Qi} or added to distort it favorably~\cite{Jach}. Given the 2.2~ms injection time required to accumulate 26~$\times$~10$^{12}$ particles at 2~mA, the addition of a 2nd-harmonic of the fundamental magnet ramp may be necessary to flatten the ramp for injection.
The Booster is commissioning an alternative approach to stabilizing the injection orbit by using the horizontal correctors to compensate for the resonant dipole ramp~\cite{Bhat}. Flattening the ramp at extraction would also facilitate matching to the RF buckets and reference momentum of the Main Injector.

If the fundamental RF frequencies for the proton complex are preserved from the PIP-II era, the RCS RF cavity design can draw from the Booster RF cavity development effort. The current Booster RF cavities are a parallel-biased iron-ferrite design and provide 60~kV of acceleration over a length of 2.25~m~\cite{Awida}. A new RF cavity design based on perpendicular-biased YAG-ferrite design may provide similar acceleration over half the length~\cite{Romanov}. A perpendicular-biased cavity is currently being commissioned for the Fermilab Booster~\cite{Madrak,Tan}. To ramp the Booster at 20~Hz at least 1.2~MV is required~\cite{PIP2}. For the Fermilab RCS, beam-loading effects will be significant and a total RF voltage of up to 2~MV may be required~\cite{Pellico}. %The RCS design would benefit from an R\&D effort in normal-conducting RF cavity design with high accelerating gradient, wide frequency sweeps, high duty factor, and robust beam-loading compensation.

The current Booster RF cavities have full apertures of 5.7~cm and there are designs of the 53 MHz RF cavities with full apertures up to 8.255~cm~\cite{Hassan}; consequently the RF cavities need not restrict the aperture of the ring even for insertions with large betas.

Currently the Fermilab Booster uses bunch rotation via quadrupole excitation to longitudinally match the Booster beam to the smaller slip-stacking buckets in the Recycler~\cite{YangBR,Ahrens}, an imperfect process which may contribute to slip-stacking losses~\cite{EldredThesis}. The 2nd-harmonic RF cavity being commissioned for the Fermilab Booster could facilitate bunch rotation~\cite{Bhat2} and during PIP-II less bunch-rotation will be required. The PIP-II operation experience with quadrupole-excitation bunch rotation will allow us to further assess this strategy for the RCS era.

A superior method for longitudinally matching to the slip-stacking buckets may be simply be an abundance of 2nd-harmonic voltage. If 2nd-harmonic RF cavities are installed on the RCS for the purpose of bunch flattening at injection (see Section~\ref{SectionLaslett}), the same method can be used for extraction.

\subsection{RCS Lattice} \label{SectionLattice}

We consider an overall RCS lattice design in the form of a series of achromatic FODO arcs each bookended by focusing triplets that convey the beam through long dispersion-free inserts. The overall lattice design can be compared to that of existing RCS and accumulator rings shown in \cite{Wei,TangRev}.

An insertion-region should be at least 6 meters in order to accommodate an H$^{-}$ injection region (e.g. \cite{PIP2}) or to accommodate four perpendicular-biased RF cavities (see Section~\ref{SectionRF}). Large beta functions are required to minimize foil heating at injection (see Section~\ref{SectionFoil}), unless laser stripping is used. The straight between the injection location and the first bending arc should have at least $2\pi/3$ phase advance to collimate particles scattered off the injection foil. The injection optics can be used in every straight section to maximize superperiodicity, or alternatively a symmetric subset of straight sections can use the injection optics. 

In a FODO cell, the peak beta function is proportional to the cell length and consequently the beta functions are minimized by the use of many alternating focusing quads. In breaking the achromatic arcs into many FODO bends, the momentum compaction factor can also be minimized to avoid transition crossing. The dipole bending radius is maximized to reduce the eddy figure of merit $\chi$ and the momentum compaction factor $\alpha_{c}$, but must fit within the desired circumference. Two or three insertion regions must be dedicated to injection, extraction, collimation, and damping; the remainder should accommodate the RF cavities.

\begin{figure}[htp]
\begin{centering}
\includegraphics[height=240pt, width=240pt]{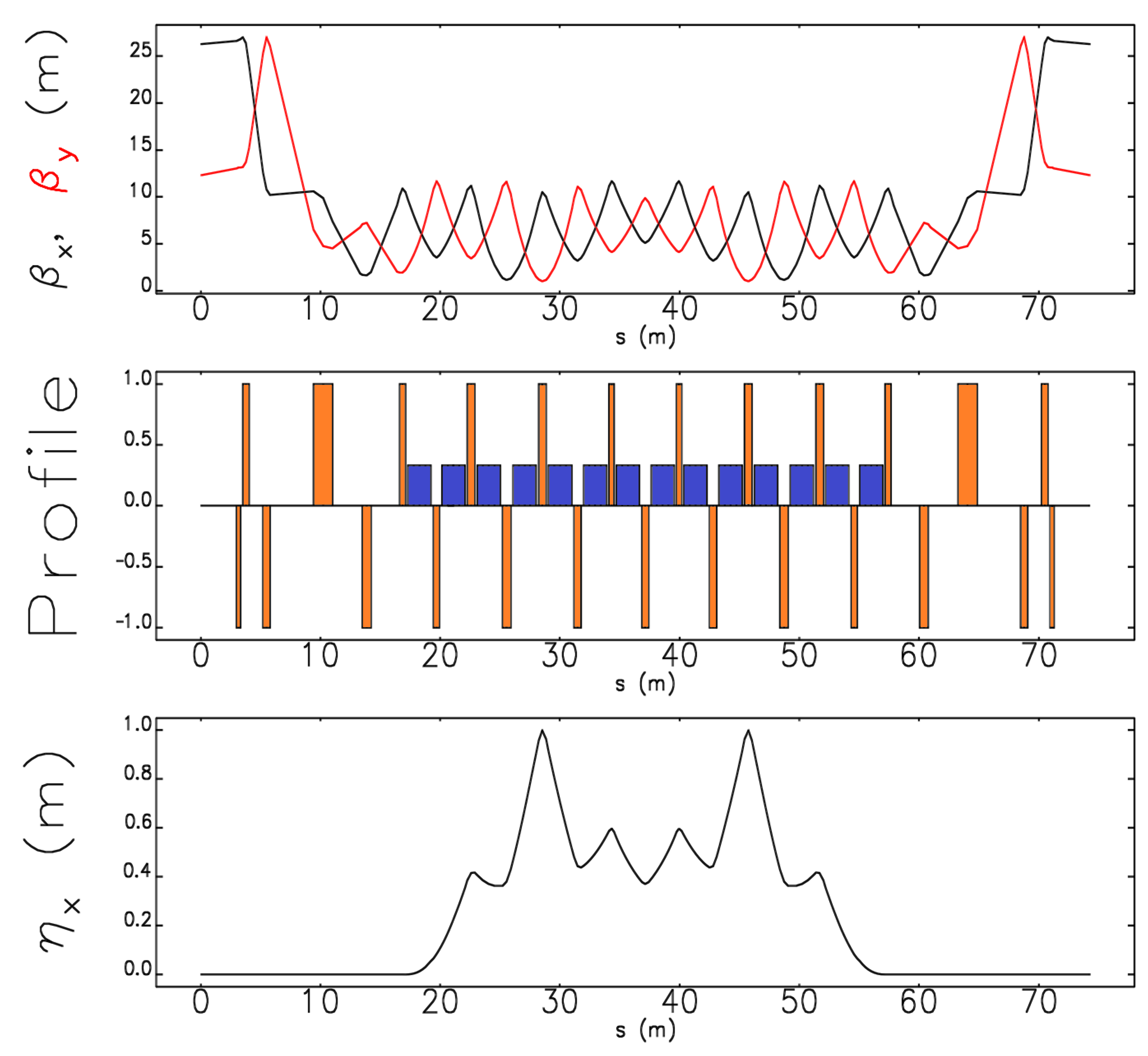}
 \caption{Twiss parameters for one of eight superperiods. (top) Horizontal and vertical beta functions shown in black and red, respectively. (middle) Location and length of magnetic lattice elements with dipoles shown as short blue rectangles and quadrupoles shown as tall orange rectangles. (bottom) Linear dispersion function.}
  \label{Lattice}
\end{centering}
\end{figure}

\begin{table}[htp]
\centering
\caption{Parameters of RCS Lattice}
\begin{tabular}{| l | c |}
\hline
Parameter & Value \\
\hline
Circumference & 600 m \\
Periodicity & 8 \\
Bend Radius & 35 m \\
Max Beta in Arc  & 13 m \\
Max Beta in Straight & 30 m \\
Max Dispersion & 1.2 m \\
Insertion Lengths & 6.5 m, 2 $\times$ 4.4 m \\
~& 2 $\times$ 2.4 m, 2 $\times$ 1.3 m \\
\hline
Betatron Tunes & 16.8,~17.9 \\
Linear Chromaticity & -24.5 \\
Transition Energy & 12.5~GeV \\
\hline
\end{tabular}
\label{Param}
\end{table}

Fig.~\ref{Lattice} shows an example lattice that meets the criteria for circumference, periodicity, injection optics, beampipe aperture, eddy figure of merit, momentum compaction factor, and total insertion length. Table~\ref{Param} shows the corresponding list of parameters. The beampipe diameter is 5~cm in the arcs and 7.6~cm in the straights sections, but the large aperture quadrupoles must be longer to maintain the same eddy figure of merit $\chi$. Sextupoles (not shown) should be interleaved into the FODO arcs at high-dispersion locations with appropriate phase-advances.

\subsection{Nonlinear Integrable Optics \& Electron Lens} \label{SectionIOTA}

The Fermilab Accelerator Science \& Technology (FAST) facility was created for advanced accelerator R\&D and beam physics research~\cite{AntipovIOTA}. Two technologies under development at FAST/IOTA, nonlinear integrable optics and electron lens space-charge compensation, can be applied to high-intensity hadron rings.

Nonlinear integrable optics is an innovation in acceleration design to provide immense nonlinear focusing without generating parametric resonances~\cite{Danilov}. Nonlinear lattices which preserve at least one invariant of motion have also demonstrated an improvement in dynamic aperture over conventional nonlinear lattice design~\cite{Ruisard,AntipovQI}.

\begin{figure}[htp]
\begin{centering}
\includegraphics[height=120pt]{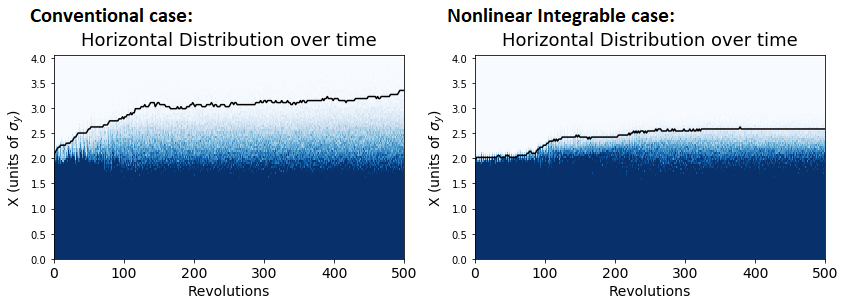}
 \caption{Horizontal particle distribution over time for conventional (left) and nonlinear integrable (right) case. The color axis is scaled to express the variation in the halo density instead of the full density range and the black line indicates the 99.9th percentile of the beam. Simulation and lattice details in \cite{Eldred17}.}
  \label{HaloX}
\end{centering}
\end{figure}

In high-intensity beams, small disturbances of the beam core cause outlying particles to be drawn into the beam halo~\cite{Wangler}. Under strong nonlinear focusing, mismatched particle beams instead rapidly reach an equilibrium distribution without generating halo~\cite{Hall,WebbArxiv}. Fig.~\ref{HaloX} shows suppressed halo formation from a 5\% mismatched Kapchinski-Vladimirski (KV) beam in a nonlinear integrable lattice~\cite{Eldred17}. The benefit of halo mitigation technology can be expressed as a reduction in uncontrolled losses and consequently an increase in loss-limited RCS intensity~\cite{ShiltsevNote}. The Landau damping provided by the nonlinear focusing can be benchmarked by an anti-damper which emulates a generic collective instability~\cite{Stern,Burov}. 

\begin{figure}[htp]
\begin{centering}
\includegraphics[height=240pt, width=240pt]{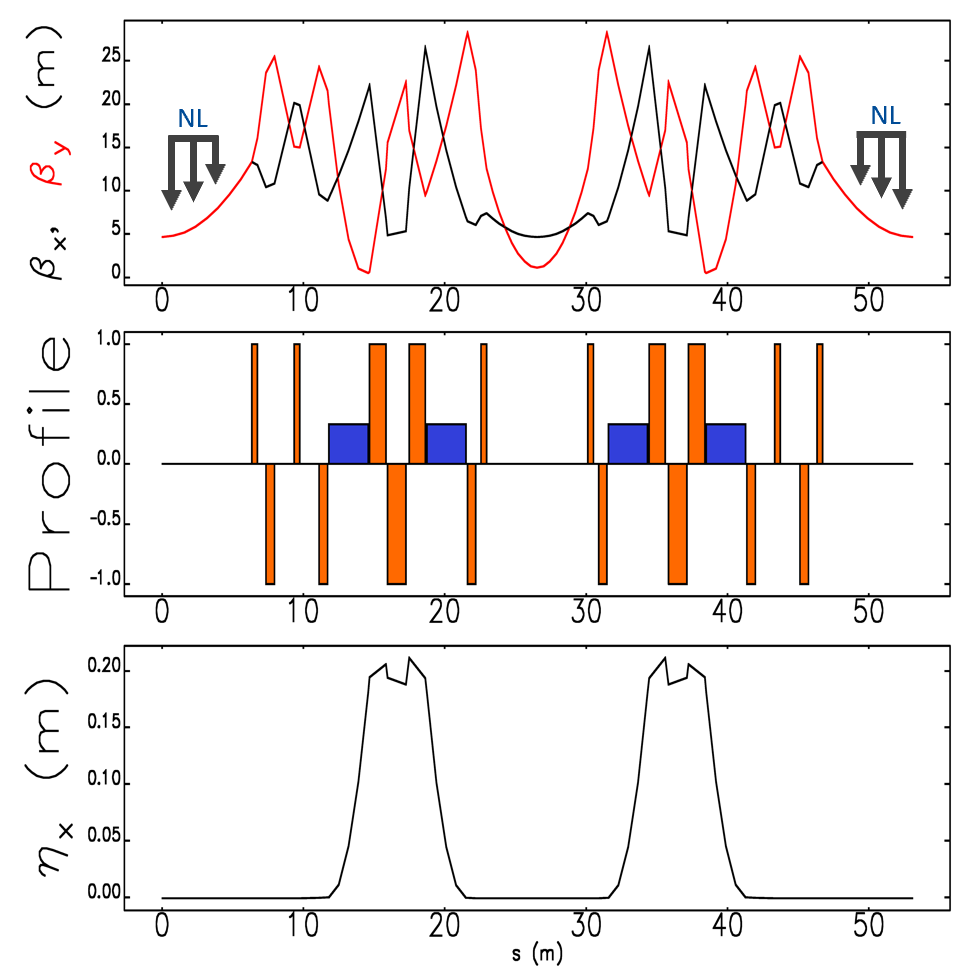}
 \caption{Twiss parameters for one of the twelve superperiods. (top) Horizontal and vertical beta functions shown in black and red, respectively. Nonlinear insertion location indicated by NL. (middle) Location and length of magnetic lattice elements with dipoles shown as short blue rectangles and quadrupoles shown as tall orange rectangles. (bottom) Linear dispersion function.}
  \label{IRCS}
\end{centering}
\end{figure}

Fig.~\ref{IRCS} shows a test lattice used to evaluate the performance of integrable RCS design in multiparticle simulations~\cite{Eldred18}. The nonlinear insertions are located in dispersion-free regions with matching beta functions and are separated by n$\pi$ phase-advances. The lattice features for nonlinear integrable optics may be reconciled with those described in Section~\ref{SectionLattice} by periodically dedicating dispersion-free drifts to nonlinear insertions and tuning the linear optics. Known technology for nonlinear insertions with integrable dynamics include precision-machined elliptic magnets and electron lenses~\cite{AntipovIOTA}.

Integrable lattice design requires precisely defined phase advances, but those phase-advanced are shifted by space-charge forces. Integrable lattices are tuned to compensate the linear space-charge tune-shift at a certain beam intensity~\cite{RomanovIOTA}. In \cite{Eldred18} a simulation of the (error-free) integrable RCS test lattice demonstrated strongly suppressed beam halo and stable nonlinear motion with a space-charge tune-shift of 0.4.

Webb et al.~\cite{Webb} shows that the chromaticities of an integrable lattice do not have to be corrected to zero; the motion of an off-momentum particle is integrable to first order if the horizontal and vertical chromaticity are matched. For a ring composed of several integrable cells the synchrotron motion is adiabatic and the beam can be stable under significant chromatic tune spread~\cite{Eldred18}.

The electron lens is a versatile particle accelerator device with applications in beam-beam compensation~\cite{ShiltsevTev}, collimation~\cite{Fitterer,Mirachi}, nonlinear focusing~\cite{ShiltsevLHC}, and space-charge compensation. In electron lens space-charge compensation, the electron beam profile is shaped to directly counteract the nonlinear space-charge defocusing effect. The electron lens is formed by an intense keV-scale electron beam co-propagating with circulating beam over a short solenoidal field~\cite{Noll}. An electron column is a variant of the electron lens in which the electron distribution is formed by ionization of a high-pressure gas by the circulated beam~\cite{Freemire}. The lattice design criteria for effective implementation of space-charge compensating electron lenses is a subject of active research.

\section{H$^{-}$ Stripping Foil Injection} \label{SectionFoil}

Protons will be accumulated in the Fermilab RCS via charge-stripping foil injection, like the PIP-II Booster. A thin carbon based foil target is used to remove two electrons from the H$^{-}$ linac beam allowing it to be combined with a circulating proton beam. The higher charge of the RCS will lead to much longer injection time than the PIP-II Booster and the effects of circulating protons scattering off the injection foil must be careful considered. We scale the foil scattering effects from a simulation of the PIP-II Booster~\cite{PIP2}, and benchmark those effects with the operational experiences of the J-PARC RCS and Oak Ridge Spallation Neutron Source (SNS).

H$^{-}$ beams can also be stripped by the use of high-power lasers, recently demonstrated at the SNS over a 10~$\mu$s timescale with 95\% stripping efficiency~\cite{Cousineau}. In this section we describe the constraints placed on an RCS facility which uses the conventional foil-stripping technique, but anticipate that laser-stripping technology may ultimately supplant it.

\subsection{Parametrics of Foil Heating}

Table~\ref{FoilT} shows a list of parameters for the stripping foil injection of the PIP-II beam~\cite{PIP2} and our RCS baseline scenario.

%\begin{table}[htp]
%\centering
%\caption{Foil Stripping Injection for PIP-II and RCS baseline scenarios}
%\begin{tabular}{| l | l | l | l | l |}
%\hline
%Description & Symbol & PIP-II & RCS & units \\
%\hline
%Injection Energy & E & 0.8 & 1.0 & GeV \\
%Foil Thickness & $\Delta x$ & 600 & 600 & $\mu$g cm$^{-2}$ \\
%Stopping Power & dE/dx & 2.1 & 2.0 &  $10^{-6}$ MeV cm$^{2}$ $\mu$g$^{-1}$ \\
%Injection Current & I & 2 & 2 & mA \\
%Ring Intensity & N & 6.6 & 26 & $10^{12}$ particle \\
%Pulse Duration & $\Delta t$ & 0.55 & 2.17 & ms \\
%95\% Norm. Emit. & $\epsilon_{n}$ & 16 & 24 & mm mrad \\
%Beta Product & $\beta_{x}~\times~\beta_{y}$ & 124 & 320 & m$^{2}$ \\
%Circ. 95\% Spot-size & $4\pi \sigma_{x}\sigma_{y}$ & 210 & 440 & mm$^{2}$ \\
%Peak Hit Density & $\rho_{h}$ & 63 & 98 & hit per particle cm$^{-2}$  \\
%Foil Heating Rate & $\Delta \text{T}_{H} / \Delta t$ & 370 & 540 & K ms$^{-1}$ \\
%\hline
%Peak Foil Temp. & T$_{max}$ & 820 & 1620 &  K \\
%\hline
%\end{tabular}
%\label{FoilT}
%\end{table}

\begin{table}[htp]
\centering
\caption{Foil Stripping Injection for PIP-II and RCS baseline scenarios}
\begin{tabular}{| l | l | l |}
\hline
Parameter & PIP-II & RCS \\
\hline
Injection Energy E & 0.8~GeV & 1.0~GeV \\
Foil Thickness $\rho \Delta x$ & 600~$\mu$g cm$^{-2}$ & 600~$\mu$g cm$^{-2}$ \\
Stopping Power $dE/ \rho dx$ & 2.1 eV cm$^{2}/\mu$g & 2.0 eV cm$^{2}/\mu$g \\
Injection Current I & 2~mA & 2~mA \\
Ring Intensity N & 6.6~$\times~10^{12}$ particles & 26~$\times~10^{12}$ particles \\
Pulse Duration $\Delta t$ & 0.55~ms & 2.17~ms \\
95\% Norm. Emit. $\epsilon_{n}$ & 16~mm mrad & 24~mm mrad \\
Beta Product $\beta_{x}~\times~\beta_{y}$ & 124~m$^{2}$ & 320~m$^{2}$ \\
Circ. 95\% Spot-size $4\pi \sigma_{x}\sigma_{y}$ & 210~mm$^{2}$ & 440~mm$^{2}$ \\
Peak Hit Density $\rho_{h}$ & 63~hits per particle cm$^{-2}$ & 98~hits per particle cm$^{-2}$ \\
Foil Heating Rate $\Delta \text{T}_{H} / \Delta t$ & 370~K ms$^{-1}$ & 540~K ms$^{-1}$ \\
\hline
Peak Foil Temp. T$_{\text{max}}$ & 830 K & 1640 K \\
\hline
\end{tabular}
\label{FoilT}
\end{table}

Given these parameters the foil heating rate is given by:
\begin{eqnarray}
\frac{dT_{H}}{dt} = \rho_{h} N \Delta t^{-1}  \frac{dE}{\rho dx} c_{p}^{-1}
\end{eqnarray}
where $dE/ \rho dx$ is the stopping power with units eV cm$^{2}/\mu$g and $c_{p}$ is the specific heat of carbon, which has a temperature dependence~\cite{Butland} characterized by:
\begin{align} \nonumber
c_{p}(T) =&~2.25374 + 3.81216 \cdot 10^{-5}~T - 377.700~T^{-1} \\
 & - 1.81792 \cdot 10^{5}~T^{-2} + 6.66549 \cdot 10^{7}~T^{-3} - 6.01191 \cdot 10^{9}~T^{-4}
\end{align}

The peak hit-density $\rho_{h}$ is the hits per particle per cm$^{2}$ at the hottest part of the foil and is specific to the painting scheme and injection time. Consider the case in which the injected beam reaches a rapid phase-space equilibrium. The hit density increases linearly with the injection turns because the hits from each circulating particle accumulate uniformly. The total hits accumulate quadratically over injection interval because the beam density increases linearly with injection time. The hit density is proportional to the beam density of the injected beam scale which is inversely proportional to the spot-size and revolution time of the circulating beam. Putting this together we can write
\begin{align}
\rho_{h}(t) &= g_{0} \Delta t \left(\epsilon_{n} \sqrt{ \beta_{x} \beta_{y}}\right)^{-1} \frac{v}{C} \left( \frac{t}{\Delta t} \right)^{2} \\
\frac{dT_{H}}{dt} &= g_{0} N \frac{dE}{\rho dx} c_{p}^{-1} \left(\epsilon_{n} \sqrt{ \beta_{x} \beta_{y}}\right)^{-1} \frac{v}{C} \left( \frac{t}{\Delta t} \right)^{2}
\end{align}
where $g_{0}$ is a geometric factor associated with the injection scheme. We can calculate $g_{0} = 0.25$ from the injection simulation of the PIP-II beam~\cite{PIP2}.

\begin{figure}[htp]
\begin{centering}
\includegraphics[height=120pt]{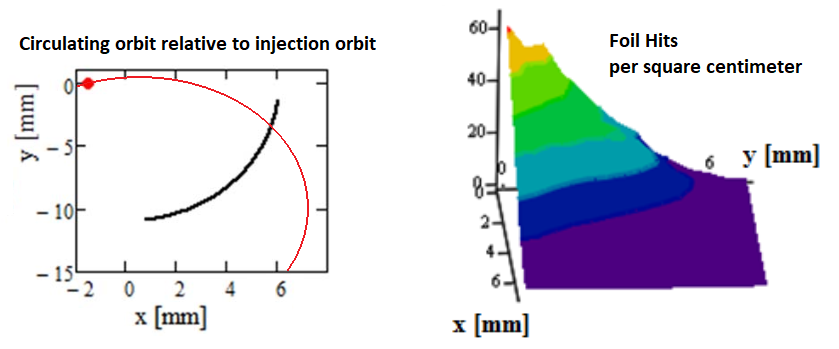}
 \caption{PIP-II injection painting and profile of foil hits~\cite{PIP2}. (left) PIP-II circulating beam orbit shown in black line, injection beam orbit shown in red point, and circulating beam size shown in red line. (right) Hits on stripping foil per cm$^{2}$ by the end of the PIP-II injection painting cycle.} 
  \label{PIP2foil}
\end{centering}
\end{figure}

Fig.~\ref{PIP2foil} shows the ``anti-correlated'' injection painting scheme for PIP-II, which follows the elliptical contour with the sum of horizontal and vertical action held constant. This can be contrasted to a ``correlated'' injection painting scheme, in which the horizontal and vertical action vary together and the injection point moves away from the core of the circulating beam. For a large ratio between the emittance of the circulating beam and the emittance of the injected beam, such as the SNS accumulator ring, correlated injection painting reduces the peak foil heating~\cite{Galambos}. In the case of the PIP-II Booster or Fermilab RCS, peak foil heating is minimized by the anti-correlated painting scheme which maintains a maximum circulating beam size at injection. Anti-correlated injection painting may also lead to more stable space-charge dynamics by reducing the peak charge-density and anticipating transverse emittance-exchange~\cite{Hotchi}. With the choice of anti-correlated injection painting for the Fermilab RCS, we scale the foil heating from the PIP-II Booster using the same geometric factor $g_{0}$.

The primary cooling mechanism for the foil is blackbody radiation:
\begin{eqnarray} \label{InjCool}
\frac{dT_{C}}{dt} = 2 e \sigma_{sb} c_{p}^{-1} (\rho \Delta x)^{-1} T^4
\end{eqnarray}
where $\rho \Delta x$ is the foil thickness in units of $\mu$g/cm$^{2}$ and where $\sigma_{sb} = 5.670~\times~10^{-8}$ W m$^{-2}$ K$^{-4}$ is the Stefan-Boltzmann constant. Conservatively we take the emissivity $e=0.5$ . The leading factor of 2 is to account for the combined surface area of both sides of the foil.

To calculate the peak temperature of the foil, one must take into account not just the cooling between injection pulses but the cooling during the injection pulse itself:
\begin{eqnarray} \label{InjHeat}
\frac{dT}{dt} = \frac{dT_{H}}{dt} - \frac{dT_{C}}{dt}  = g_{0} N \frac{dE}{\rho dx} c_{p}^{-1} \left(\epsilon_{n} \sqrt{ \beta_{x} \beta_{y}}\right)^{-1} \frac{v}{C} \left( \frac{t}{\Delta t} \right)^{2} - \sigma_{sb} c_{p}^{-1} (\rho \Delta x)^{-1} T^4
\end{eqnarray}

The peak temperature can be obtained by integrating Eq.~\ref{InjHeat} and Eq.~\ref{InjCool} in an alternating fashion according to the injection pulse structure. Since most of the cooling occurs immediately after injection, the peak temperature is not very sensitive to the injection repetition rate.

\subsection{Foil Injection Scenarios}

At temperatures above $\sim$1800 K, the carbon graphite foil sublimates and this is a major factor in the lifetime of the foil. The change in foil thickness is given by
\begin{align}
 \frac{dh(t)}{dt} \frac{1}{\rho} = - 8.12 \times 10^{10} \frac{[\text{g}\sqrt{\text{K}}]}{[\text{cm}^{2} \text{s}]} \frac{\exp\left( - \frac{83500}{T} \right)}{T^{1/2}}.
\end{align}

From the exponential dependence on temperature in this formula~\cite{LebedevFoil} we can see that the peak temperature at the end of the injection pulse dominates the removal of foil material~\cite{Blokland}. At a peak temperature of 1950 K, the foil will last about a year of RCS operation before removing 50 $\mu \text{g}/\text{cm}^{2}$. At a peak temperature of 2150 K, the foil will last about a week before encountering the same benchmark. Hybrid-Boron Carbon (HBC) foils have been shown to achieve superior foil-lifetimes at high temperatures~\cite{Sugai1,Sugai2,Yoshimoto}.

\begin{figure}[htp]
\begin{centering}
\includegraphics[height=180pt, width=360pt]{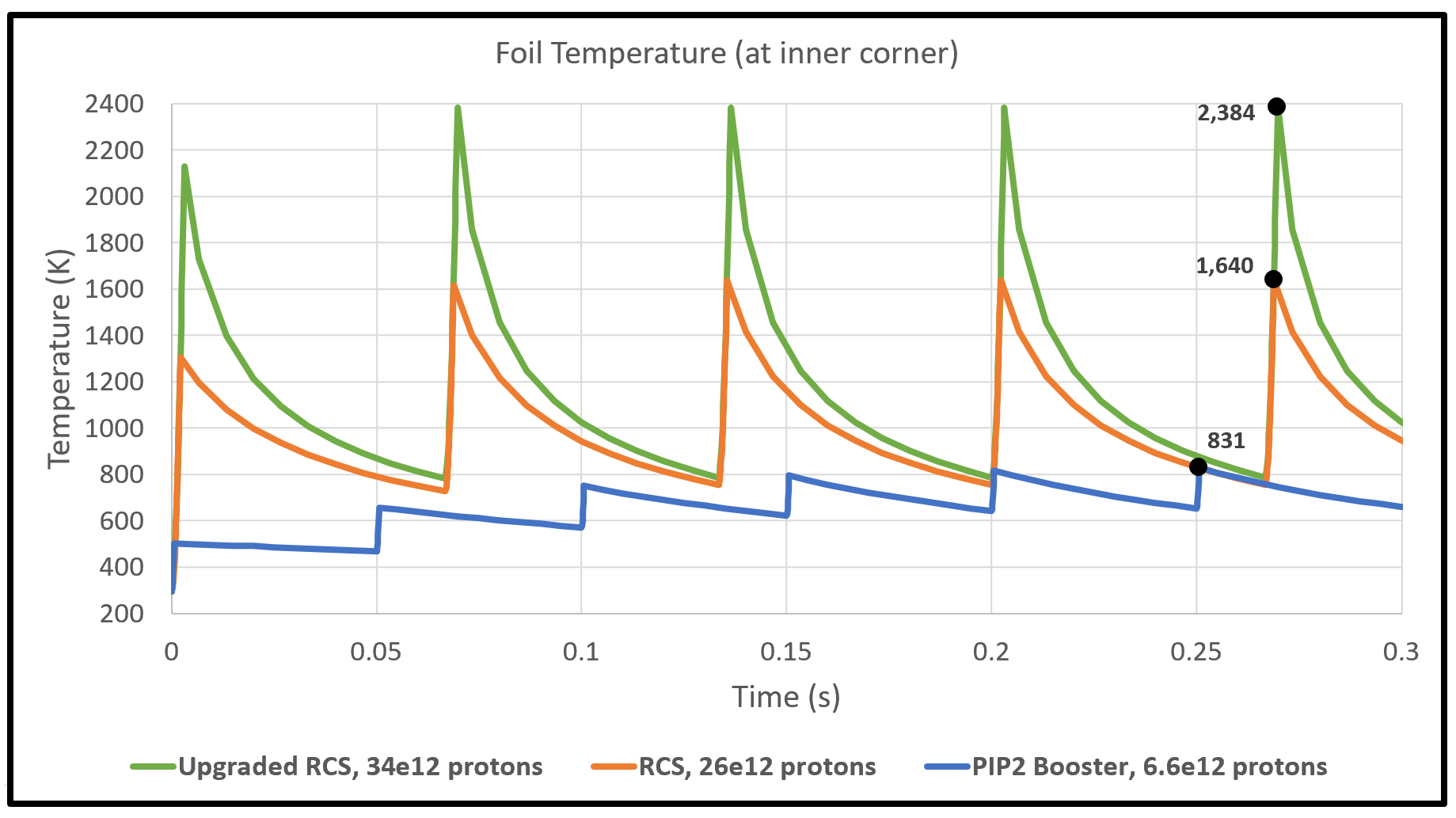}
 \caption{Foil temperature at hottest part of carbon foil for PIP-II Booster, proposed RCS scenario, and a higher intensity upgrade of the RCS.}
  \label{FoilF}
\end{centering}
\end{figure}

Considering the RCS baseline scenario, we set the intensity $N$ to 26 $\times~10^{12}$ protons, the emittance $\epsilon_{N}$ to 24~mm~mrad, the beta product to 320 m$^{2}$, the foil thickness to 600 $\mu \text{g}/\text{cm}^{2}$, and the circumference to 600 m. These parameters are summarized in Table~\ref{FoilT}. Numerically integrating with these parameters, we obtain a peak temperature of $\sim$1630~K as shown in Fig.~\ref{FoilF}. At this peak foil temperature, the carbon foil lifetime is very manageable. 

Fig.~\ref{FoilF} also shows a higher intensity upgrade of the RCS which would encounter a severe foil heating limit. If foil heating becomes a limiting factor in RCS intensity, one strategy would be to fill only part of the RCS circumference and accumulate beam in the Main Injector with a larger number of partial batches. Fig.~\ref{Batch} shows the number of batches that can be accumulated in the Main Injector via slip-stacking for other batch-lengths.

\subsection{Foil Scattering \& Shielding}

The thickness of the carbon injection stripping foil is determined by the stripping efficiency, which should be at least 99\%. For the PIP-II Booster the injection foil needs to be at least 380 $\mu$g/cm$^{2}$ to achieve the 99\% stripping efficiency benchmark, but an injection foil thickness of 600 $\mu$g/cm$^{2}$ is chosen to provide a longer foil lifetime at greater stripping efficiency~\cite{PIP2}. At higher injection energies, the required foil thickness for the same stripping efficiency scales as $\beta^{2}$~\cite{Chou}. For the Fermilab RCS, we assume continued operation with the 600 $\mu$g/cm$^{2}$ foil thickness, with an almost 200 $\mu$g/cm$^{2}$ margin on 99\% stripping efficency. The particles not completely stripped by passage through the injection foil are overwhelmingly excited H$^{0}$ ions. 

Exposure of the injection foil to the circulating beam is minimized by injecting near to the edge of the injection foil. Consequently, there is approximately 1\% of the injected H$^{-}$ particles which completely miss the injection foil.

The less than 2\% of injected particles which either miss the primary foil or are not completely stripped by the foil have an average beam power of about 1kW. If the injection foil is placed in the center of a four-bump orbit of the RCS, these particles will be displaced away from the RCS orbit by the downstream inner bump corrector and can directed be to a high-power beam dump~\cite{Saha}. These ion beams can also be directed to thicker secondary foils to strip the electrons and facilitate transfer to the injection dump.

The scattering of circulating protons of the injection foil is also an important source of particle loss. At both SNS and J-PARC, the overwhelmingly highest radioactivation levels are the injection region and the area immediately downstream from injection~\cite{Plum,Yamamoto}. Scattering losses can be expressed in the terms of foil hits, foil thickness, and the Coulomb scattering collision-length~\cite{Macek}. The Coulomb scattering collision-length is a steep function of acceptance angles $\theta_{x}$ and $\theta_{y}$ at the injection location
\begin{align}
\lambda = \left( \frac{2 e^{2}}{\gamma \beta^{2} m_{p} c^{2}} \right)^{2} \left( \frac{Z^{2}}{A} \right) \left[ \frac{1}{\theta_{x} \theta_{y}} + \frac{1}{\theta_{x}^{2}} \text{atan} \left( \frac{\theta_{y}}{\theta_{x}} \right) + \frac{1}{\theta_{y}^{2}} \text{atan} \left( \frac{\theta_{x}}{\theta_{y}} \right) \right]
\end{align}

If we use the Fermilab RCS parameters given in Table~\ref{FoilT} to determine the acceptance angles, the foil scattering rate is $\sim 0.7\%$ for 420 W of radiated power. Comparing this foil scattering loss to the $\sim 70$ W scattered at SNS, the Fermilab RCS losses need careful management.

Particles with extreme divergence at the injection foil will be lost at beampipe apertures downstream as the betatron phase advances. The set of focusing quads and bump correctors immediately downstream of the injection foil should have significant shielding and use a larger aperture. Two-stage secondary collimators should cover $2\pi/3$ of phase-advance between the injection foil and the first bending arc. This downstream injection scattering collimator is a feature the SNS accumulator ring lacks, which instead relies on its main collimators in the straight section after the first bending arc. The J-PARC RCS places its main collimators between the injection region and the first bending arc. For the Fermilab RCS we suggest dedicated scattering collimators before the first bending arc, followed by the main collimation section after the arc. The feature of significant phase-advance within the injection straights should be considered in the RCS lattice design. Booster experience~\cite{Kapin} indicates collimator length is also a critical factor in containing secondary scattering particles.

\section{Beam Power Upgrades Leveraging RCS} \label{SectionUpgrades}

The 2.4~MW upgrade of the Fermilab proton complex should be compatible with further improvement of the accelerator program. As Fig.~\ref{History} indicates, there has been a sustained pattern of increasing demand for Main Injector beam power for the long-baseline neutrino program. Table~\ref{Upgrades} summarizes some upgrade options for 3+~MW which are described in detail in this section. For all scenarios, we assume that appropriate high-power target stations would be available. The ultimate power limits of the Main Injector itself must still be determined.

%\begin{table}[htp]
%\centering
%\caption{Comparison between several high-power upgrades of RCS-based facility}
%\begin{tabular}{| l | l | l | l | l | l |}
%\hline
%Acc. Ring & SR & MI & MI & MI & SR \\
%Acc. Technique & Slip-stacking & Boxcar & Slip-stacking & Slip-stacking & Slip-stacking \\
%Inj. Energy & 1~GeV & 2~GeV & 2~GeV & 1~GeV & 1~GeV \\
%MI ramp time & 1.2~s & 1.2~s & 1.2~s & 0.65~s & 0.65~s \\ 
%\hline \hline
%RCS Intensity & 25~$\times~10^{12}$ & 55~$\times~10^{12}$ & 34~$\times~10^{12}$ & 25~$\times~10^{12}$ & 25~$\times~10^{12}$ \\
%Batch Total & 4+5 & 5 & 5+5 & 4+5 & 4+5 \\
%MI Intensity & 225~$\times~10^{12}$ & 275~$\times~10^{12}$ & 340~$\times~10^{12}$ & 225~$\times~10^{12}$ & 225~$\times~10^{12}$ \\
%MI cycle time & 1.200~s & 1.600~s & 1.867~s & 1.267~s & 0.650~s \\
%\hline \hline
%MI Beam Power & 3.6~MW & 3.3~MW & 3.5~MW & 3.4~MW & 6.6~MW \\
%\hline
%\end{tabular}
%\label{Upgrades}
%\end{table}

\begin{table}[htp]
\centering
\caption{Comparison between several high-power upgrades of RCS-based facility}
\begin{tabular}{| l | l | l | l | l |}
\hline
Scenario & Baseline & Linac & Linac & MI Ramp \\
\hline
\quad Accum. Technique & Slip-stacking & Boxcar & Slip-stacking & Slip-stacking \\
\quad Inj. Energy & 1~GeV & 2~GeV & 2~GeV & 1~GeV \\
\quad MI ramp time & 1.2~s & 1.2~s & 1.2~s & 0.65~s \\ 
\quad RCS Intensity & 25~$\times~10^{12}$ & 55~$\times~10^{12}$ & 34~$\times~10^{12}$ & 25~$\times~10^{12}$ \\
\quad Batch Total & 4+5 & 5 & 5+5 & 4+5 \\
\quad MI Intensity & 225~$\times~10^{12}$ & 275~$\times~10^{12}$ & 340~$\times~10^{12}$ & 225~$\times~10^{12}$ \\
\hline 
MI Accumulation & ~ & ~ & ~ & ~ \\
\hline
\quad MI cycle time & 1.800~s & 1.600~s & 1.867~s & 1.267~s \\
\quad MI Beam Power & 2.4~MW & 3.3~MW & 3.5~MW & 3.4~MW \\
\hline
SR Accumulation & ~ & ~ & ~ & ~\\
\hline
\quad MI cycle time & 1.200~s & 1.200~s & 1.200~s & 0.650~s \\
\quad MI Beam Power & 3.6~MW & 4.4~MW & 5.4~MW & 6.6~MW \\
\hline
\end{tabular}
\label{Upgrades}
\end{table}

Slip-stacking currently takes place in the Fermilab Recycler, so the long beam accumulation time does not contribute to the Main Injector cycle time. In this paper we have laid out a scenario for 2.4~MW which is a continuation of slip-stacking but that does not benefit from the Fermilab Recycler as an accumulator. An 11~GeV storage ring (SR) as a functional replacement of the Fermilab Recycler would be a compelling path to 3.6~MW by reducing the Main Injector cycle time from 1.8~s to 1.2~s. The SR could improve on the Recycler design by using separate function magnets, feature a larger aperture, and precisely match the Main Injector circumference. The SR would physically take the place of the Recycler in the Main Injector tunnel and the total cost of the permanent magnets would be very modest compared to other upgrade options.

If the SR was constructed in tandem with the Fermilab RCS, slip-stacking accumulation would not have to be commissioned in Main Injector. The facility would reach 2~MW when five batches of 25$~\times~10^{12}$ protons were conventionally accumulated in the SR and beam power would rise steadily as slip-stacking was commissioned.

The Project X ICD-2~\cite{ICD2} envisions an RCS-based high-power Main Injector program in combination with a 2~GeV linac-based kaon and muon program. At that injection energy, the RCS could accumulate as many as 55~$\times~10^{12}$ protons before reaching a Laslett tune-shift of 0.3. The corresponding injection time would rise to 4.76~ms or 0.14$\pi$ into the 15-Hz ramp-cycle and a reduction of the ramp rate to 10~Hz may be necessary. Fig.~\ref{FoilF} also indicates a foil heating limit would be reached around 34~$\times~10^{12}$ protons unless a laser-stripping program could be implemented. With the linac current restricted to 2~mA, any increase in linac energy above 2~GeV may not provide any clear improvement in RCS power.

For a 10~Hz RCS with 55~$\times~10^{12}$ proton batches accumulated via conventional boxcar stacking, the Main Injector intensity is 275~$\times~10^{12}$ proton and the corresponding beam power is 3.3~MW. If slip-stacking is used, the beam power is clearly limited by the maximum attainable intensity of the Main Injector. Ten batches, each 550~m in length with 34~$\times~10^{12}$ protons, accumulated via slip-stacking would correspond to 3.5~MW beam power and 340~$\times~10^{12}$ protons in the Main Injector.

The Proton Driver Study II evaluated the possibility of shortening the Main Injector ramp by nearly a factor of two by doubling the Main Injector magnet power-supply voltage~\cite{PDriver}. The contemporary application to the RCS-based accelerator facility was assessed by Kourbanis in \cite{Kourbanis}. With the slip-stacking RCS facility described in this paper, a 1.267~s Main Injector cycle time could be achieved to provide 3.5~MW beam power in the Main Injector. Combining the shortened Main Injector ramp with the SR for accumulation, the Main Injector may be further decreased to 0.65~s for an extraordinary potential beam power of 6.6~MW. The Main Injector intensity would not increase in this case but as a result of the high-rep rate the realizable beam power would likely be limited by the Main Injector loss budget.

\section{Outlook}

We provide a generalized design of an RCS-based proton facility to support 2.4~MW beam power for the DUNE/LBNF long-baseline neutrino program. We perform a parametric analysis of critical design choices such as RCS circumference, intensity, beampipe aperture, and injection optics. In particular, we find an increase in the RCS energy above 8~GeV is an effective way to support low-loss operation of the high-power Main Injector. The design analysis demonstrates the continuing value of slip-stacking accumulation for high-power operation and we address relevant constraints on the technique. We show that replacing the Recycler with a new storage ring is a cost-effective method to provide a margin of safety on 2.4~MW operation and potentially enable 3.6~MW operation. We highlight the importance of lattice superperioidicy, resistive beampipe, 2nd-harmonic RF cavities, IOTA technologies, and injection scattering collimation.

Several R\&D topics remain to be addressed. The development of high-power neutrino targets will be essential part of the Fermilab LBNF program as well as the increase in beam power beyond 2.4~MW. The ultimate limits of space-charge and beam intensity in the Main Injector should be more rigorously established by empirical and theoretical studies. The Fermilab Booster can also serve as an important case study on periodic lattice design. Tests of integrable optics and electron lens technology will be performed at FAST/IOTA over the next several years. Hardware components of the RCS accelerator, such as dipole magnets and RF cavities, should undergo a renewed development effort to bring about a fully optimized and realizable accelerator facility. The user program for the hundreds of kilowatts available at 11~GeV must still be established.

\section*{Acknowledgments}

Operated by Fermi Research Alliance, LLC under Contract No. DE-AC02-07CH11359 with the United States Department of Energy.

\end{document}